\documentclass[fleqn,usenatbib]{mnras}

\usepackage{newtxtext,newtxmath}

\usepackage[T1]{fontenc}
\usepackage{ae,aecompl}

\usepackage{graphicx}	
\usepackage{amsmath}	
\usepackage{amssymb}	
\usepackage{color}
\usepackage{soul}

\newcommand{\tR}[1]{\textcolor[rgb]{0.8,0.0,0.0}{#1}}

\newcommand{\gppr}{\stackrel{>}{\scriptstyle \sim}}
\newcommand{\gappr}{\raisebox{-0.4ex}{$\gppr$}}
\newcommand{\lppr}{\stackrel{<}{\scriptstyle \sim}}
\newcommand{\lappr}{\raisebox{-0.4ex}{$\lppr$}}

\title[A WDs catalogue from {\it Gaia}-DR2 and the VO]{A White Dwarf catalogue from {\it Gaia}-DR2 and the Virtual Observatory}

\author[F. M. Jim\'enez-Esteban et al.]{
F. M. Jim\'enez-Esteban$^{1,2,3}$\thanks{E-mail: fran.jimenez-esteban@cab.inta-csic.es}, S. Torres$^{4,5}$, A. Rebassa-Mansergas$^{4,5}$,  
\newauthor
G. Skorobogatov$^{4}$, E. Solano$^{1,2}$, C. Cantero$^{4}$, C. Rodrigo$^{1,2}$
\\
\\
$^{1}$Departmento de Astrof\'{\i}sica, Centro de Astrobiolog\'{\i}a (CSIC-INTA), ESAC Campus, Camino Bajo del Castillo s/n,\\
E-28692 Villanueva de la Ca\~nada, Madrid, Spain\\
$^{2}$Spanish Virtual Observatory, Spain\\
$^{3}$ Suffolk University, Madrid Campus, C/ de la Vi\~na 3, 28003, Madrid, Spain\\
$^{4}$ Departament de F\'{\i}sica, Universitat Polit\'{e}cnica de Catalunya, c/Esteve Terrades 5, 08860 Castelldefels, Spain\\
$^{5}$ Institut d'Estudis Espacials de Catalunya, Ed. Nexus-201, c/Gran Capit\'a 2-4, 08034 Barcelona, Spain
}

\date{Accepted 2018 July 16. Received 2018 June 28; in original form 2018 May 20}

\pubyear{}

\begin{document}
\label{firstpage}
\pagerange{\pageref{firstpage}--\pageref{lastpage}}
\maketitle

\begin{abstract}
We present a catalogue of 73,221 white dwarf candidates extracted from the astrometric and photometric data of the recently published {\it Gaia} DR2 catalogue. White dwarfs were selected from the {\it Gaia} Hertzsprung-Russell diagram with the aid of the most updated population synthesis simulator. Our analysis shows that {\it Gaia} has virtually identified all white dwarfs within 100\,pc from the Sun. Hence, our sub-population of 8,555 white dwarfs within this distance limit and the colour range considered, $-\,0.52<(G_{\rm BP}-G_{\rm RP})<0.80$, is the largest and most complete volume-limited sample of such objects to date. From this sub-sample we identified 8,343 CO-core and 212 ONe-core white dwarf candidates and derived a white dwarf space density of $4.9\pm0.4\times10^{-3}\,{\rm pc^{-3}}$. A bifurcation in the Hertzsprung-Russell diagram for these sources, which our models do not predict, is clearly visible. We used the Virtual Observatory tool VOSA to derive effective temperatures and luminosities for our sources by fitting their spectral energy distributions, that we built from the UV to the NIR using publicly available photometry through the Virtual Observatory. From these parameters, we derived the white dwarf radii. Interpolating the radii and effective temperatures in hydrogen-rich white dwarf cooling sequences, we derived the surface gravities and masses. The {\it Gaia} 100\,pc white dwarf population is clearly dominated by cool ($\sim$\,8,000 K) objects and reveals a significant population of massive ($M \sim 0.8 M_{\odot}$) white dwarfs, of which no more than $\sim$\,$30-40\,\%$ can be attributed to hydrogen-deficient atmospheres, and whose origin remains uncertain. 
\end{abstract}

\begin{keywords}
white dwarfs -- stars: evolution -- Galaxy: stellar content-- astronomical data bases: miscellaneous -- catalogues -- virtual observatory tools 
\end{keywords}



\section{Introduction}

White dwarfs are the evolutionary end-point of low- and intermediate-mass stars, namely, those  with masses at the main sequence $\leq10\,\pm\,2\,M_{\sun}$. Thus, white dwarfs are one of the most common stellar objects. Their structure is relatively simple and their evolutionary properties are reasonably well understood (see the review by \citealt{Althaus2010a} and reference therein for a thorough  discussion of this issue). White dwarfs contain a core of degenerate matter. Massive white dwarfs, those with $M\,\gappr\,1.04\,M_{\sun}$, (e.g., \citealt{Althaus2007,GBerro1994})
have an oxygen and neon (ONe) core; the core of those less massive than $\sim0.45\,M_{\sun}$ is made up of helium (He) \citep[e.g.][]{Serenelli2001}; and those with masses in between, which are the most of them, have a carbon and oxygen (CO) core \citep[e.g.][and references therein]{Althaus2010a}. Given that the life-times of the progenitor stars of He-core white dwarfs exceed the Hubble time, the observed He-core white dwarf population is expected to be the product of binary star evolution where episodes of mass transfer occurred \citep[e.g.][]{Rebassa11}.

The bulk of the mass of a white dwarf is concentrated in the degenerate core acting as an energy reservoir. Regardless of their core composition, white dwarfs are surrounded by an extremely thin atmospheric layer. The envelope, which approximately represents only 0.01\% of the total mass of the star, is the responsible for the energy outflow. This envelope is basically formed by a He thin layer with a mass ranging from  $10^{-4}\,M_{\sun}$ to $10^{-2}\,M_{\sun}$. Larger envelope masses trigger He ignition at the base of the shell. The He layer is, in theses cases, surrounded by an even thinner layer of hydrogen with masses in the range $10^{-15}\,M_{\sun}$ to  $10^{-4}\,M_{\sun}$ \citep[e.g.][and references therein]{Bradley1998,Rohrmann2001,Castanheira2008,Tremblay2008}. For approximately 20\% of white dwarfs this hydrogen envelope is lost \citep{Eisenstein2006,Bergeron2011,Koester2015}. From a phenomenological point of view, white dwarfs that exhibit hydrogen lines in their spectra are named as DA, while the absence of this feature is generically referred to as non--DA white dwarfs. In particular, those which only present He features are known as DB white dwarfs. 

White dwarfs are degenerate and long living objects, i.e., thermonuclear reactions have ceased and their evolution is a long gravothermal cooling process only supported  by the  pressure of degenerate electrons. Simple theoretical considerations show that the cooling timescales of these stars are very long, $\approx10\,$Gyr. Consequently, the white dwarf population retains important information of the history of the Galaxy. Much of this information can be recovered by analysing the white dwarf luminosity function, which is defined as the number of white dwarfs per bolometric magnitude unit and cubic parsec. First derived by \citet{weid68} four decades ago, the white dwarf luminosity function has been used since then as a valuable tool to understand the nature and the history of the  different components of our Galaxy \citep[see][for a comprehensive and recent review]{GBerroOswalt2016}. For instance, the white dwarf luminosity function has been used in the study of both the thin and thick discs \citep{Winget87,GB88b,GBerro1999,Torres2002,Rowell2013}, the halo \citep{Mochkovitch1990,Isern1998,GBerro2004,vanOirschot2014} and more recently, the bulge \citep{Calamida2014,Torres2018}. 

The white dwarf mass function has also been intensively studied during the last decades. These studies show a clear concentration of white dwarfs at $\sim$0.6\,M$_{\odot}$ \citep[e.g.][]{Koesteretal79,Holbergetal08,Kepleretal15}, as well as the existence of a low-mass peak at $\sim$0.4\,M$_{\odot}$ \citep{Liebert05,Girvenetal11,Kleinmanetal13}. As mentioned before, the existence of such low-mass He-core white dwarfs is mainly due to mass transfer interactions in binaries \citep{Rebassa11,Kilicetal12}. Moreover, additional observational studies suggest the existence of an excess of massive white dwarfs near 1\,M$_{\odot}$ of unclear origin \citep{Liebert05,Giammichele12,Rebassa15-1,Rebassa15-2}.

White dwarfs have also been employed in the study and characterisation of open and globular clusters. For instance, the age of clusters, subpopulation identification, and white dwarf cooling, are only a few of the most representative examples which have been studied by \cite{Salaris2001},  \cite{Calamida2008}, \cite{GBerro2010}, \cite{Jeffery2011}, \cite{Hansen2013}, and \cite{Torres2015}, among several others. Additionally, the white dwarf population has been used as an astrophysical laboratory to test non-standard physical theories, such as the variability of the gravitational constant $G$ \citep{GBerro1995,GBerro2011}, or the existence of exotic particles \citep{Isern1992,Isern2008,Dreiner2013,Miller2014}. 

One of the major drawbacks that hampered some of the previously mentioned studies, specially those performed a few decades ago, was the scarcity and incompleteness of the samples under study. Fortunately, the advent of large-scale automatic surveys, like the SuperCOSMOS Sky Survey \citep{Hambly1998} or the Sloan Digital Sky Survey \citep{Yorketal2000} --- to cite just two representative examples ---,  have substantially increased the number of known white dwarfs. However, the lack of accurate distances, joint to the bias effects derived from selection cuts in these magnitude-limited samples, have hampered the achievement of more concluding results. In this sense, the superb astrometric capabilities of {\it Gaia} is expected to provide us with an unprecedented number of white dwarfs with excellent astrometric measurements. In particular, \cite{Torres2005}, with the aid of population synthesis models, showed that the expected number of {\it Gaia} white dwarfs will be close to 12,000 up to 100\,pc, and nearly 400,000 up to 400 pc. Moreover, the same study showed that the white dwarf population accessible by {\it Gaia} should be nearly complete up to 100\,pc and with parallax errors smaller than 10 per cent. Thus, the use of the expected superb {\it Gaia} data will open an unprecedented range of possibilities in the study of the physics of white dwarfs and the structure and evolution of the Galaxy  \citep{Barstow2014}.

However, not only identifying {\it Gaia} white dwarfs, but also deriving their stellar parameters is required to fully characterise the observed population. In this sense, many multi-wavelength photometric deep surveys are currently available, and the spectral energy distribution (SED) of the celestial objects, including faint white dwarfs, can be built with unprecedented coverage. The Virtual Observatory (VO) provides easy and fast access to these catalogues. Furthermore, the new VO tools permit the study of thousands of objects at once that any other way would be unthinkable. Thus, the VO provides us with the ideal framework to carry out the present study.

In this paper we present a simple method to identify white dwarf candidates in the {\it Gaia}-DR2 catalogue, and we use the VO to characterize the selected sample. The paper is organized as follows: in Section \ref{method} we describe the methodology used to search for white dwarf candidates; we present the catalogue of {\it Gaia}-selected white dwarfs in Section \ref{results}; the creation of the SEDs and their fits are explained in Section \ref{vosa}; we study the physical properties of the candidate white dwarfs in Section \ref{phys-prop}; and finally, we present our conclusions in Section \ref{conclusion}.


\section{Search methodology}
\label{method}

In this section we describe our methodology for identifying white dwarf candidates within {\it Gaia} DR2. We begin describing our Monte Carlo code developed to reproduce the synthetic population of {\it Gaia} white dwarfs. Based on the location of the synthetic data in the HR diagram, we then define our target selection criteria for such low-luminosity objects.

\subsection{Population synthesis code}
\label{s-popsyn}

Our population synthesis code, based on Monte Carlo techniques, has been widely used in the study of the white dwarf population of the different components of the Galaxy, as well as in globular and open clusters \citep[and references therein]{GBerro1999, GBerro2004, GBerro2010, Torres2001, Torres2002, Torres2015, Torres2016}. In particular, a comprehensive study of the capabilities of {\it Gaia} and previous estimates of the Galactic white dwarf population can be found  in \cite{Torres2015}. In what follows we describe the main ingredients employed in our simulations.

The heart of any Monte Carlo simulator is a random number generator that provides a uniform probability density and ensures a large repetition period exempt of correlations. In our case, we used the random number generator of \cite{James1990}, which guarantees excellent statistical properties and has a repetition period $\ge10^{18}$, virtually infinite for our purposes. 

The mass of main-sequence stars was randomly chosen according to a Salpeter-like initial mass function with the standard slope, $\alpha=-2.35$ \citep{Salpeter1955}. Then, three types of white dwarf populations were considered according to the components of the Galaxy, thin and thick disc, and halo. For the thin disc population, we adopted an age of 9.2\,Gyr with a constant star formation rate, while the spatial distribution of the synthetic stars followed a double exponential profile with a scale height of 250\,pc and a scale length of 2.6\,kpc, in accordance with \citet{Torres2016}. The thick disc was modeled from a star formation rate peaked at 10\,Gyr in the past and extended up to 12\,Gyr. Similarly to the thin disc, the thick disc population followed a double exponential spatial distribution but with a scale height of 1,500\,pc and a scale length of 3.5\,kpc, in agreement with  \cite{Castellani2002} and \cite{Bland-Hawthorn2016}. Finally, the born time for synthetic stars of the halo population was randomly assigned within a burst of constant star formation lasting 1 Gyr that happened 13.5\,Gyr in the past. Besides, halo stars were located according to an isothermal sphere density with a 5\,kpc core radius and adopting a typical 8.5\,kpc galactocentric distance to the Sun (see \citealt{Cojocaru2015} and references therein).

Once the born time and the mass of our synthetic stars were generated, we evaluated which of these stars had had time to evolve to white dwarf. We used an updated set of white dwarf evolutionary cooling sequences, which self-consistently include the full evolution of their progenitor stars, starting at the Zero Age Main Sequence, all the way through the central hydrogen and helium burning, the thermally-pulsing AGB, and the post-AGB phases. These sequences are metallicity dependent and encompass the full range of white dwarf masses (from 0.5 up to 1.04\,$\rm{M_{\odot}}$ for CO-core white dwarfs and above this value up to the Chandrasekhar mass for ONe-core white dwarfs). In particular, we adopted a solar metallicity value for thin and thick disc stars and a sub-solar value for halo stars. Additionally, for each of the white dwarfs generated, we considered hydrogen-rich or hydrogen-deficient atmospheres, i.e. DA or non--DA white dwarfs, respectively. For each white dwarf we randomly chose its spectral type according to the canonical distribution of $80\%$ DA and $20\%$ non--DA \citep{Eisenstein2006,Bergeron2011,Koester2015}, regardless of its effective temperature. Specifically, for CO-core white dwarfs ($M<1.04\,M_{\odot}$) and with hydrogen-rich atmospheres, we used the cooling sequences of  \cite{Renedo2010} for solar metallicity, and those of \cite{Althaus2015} for sub-solar metallicity. White dwarfs with hydrogen-deficient atmospheres and CO-cores were modeled following \cite{Camisasa2017}. Finally, for white dwarf masses larger  than $1.04\,M_{\sun}$, we used  the evolutionary sequences for DA ONe white dwarfs of \cite{Althaus2005} and \cite{Althaus2007}. Colours and magnitudes were then interpolated in the corresponding cooling sequences and transformed from the Johnson-Cousins system to the {\it Gaia} photometric system following \cite{Jordi2010}.

The simulated synthetic population of white dwarfs for the three Galactic components was then mixed proportionally to 80:15:5 for the thin disc, thick disc, and halo, respectively, in agreement with \cite{Rowell2011}, and normalized to the local space density of white dwarfs of $4.8\times 10^{-3}\,{\rm pc^{-3}}$ as estimated by \cite{Holberg2016}.

Finally, and in order to mimic the observational procedure, we introduced a photometric and an astrometric error for each of our simulated objects as prescribed in the {\it Gaia} performances\footnote{http://www.cosmos.esa.int/web/gaia/science-performance} and outlined in \cite{Bruijne2005}.


\subsection{Definition of the search criteria}
\label{s-criteria}

\begin{figure}
\includegraphics[trim=18mm 35mm 18mm 37mm,clip,width=\columnwidth]{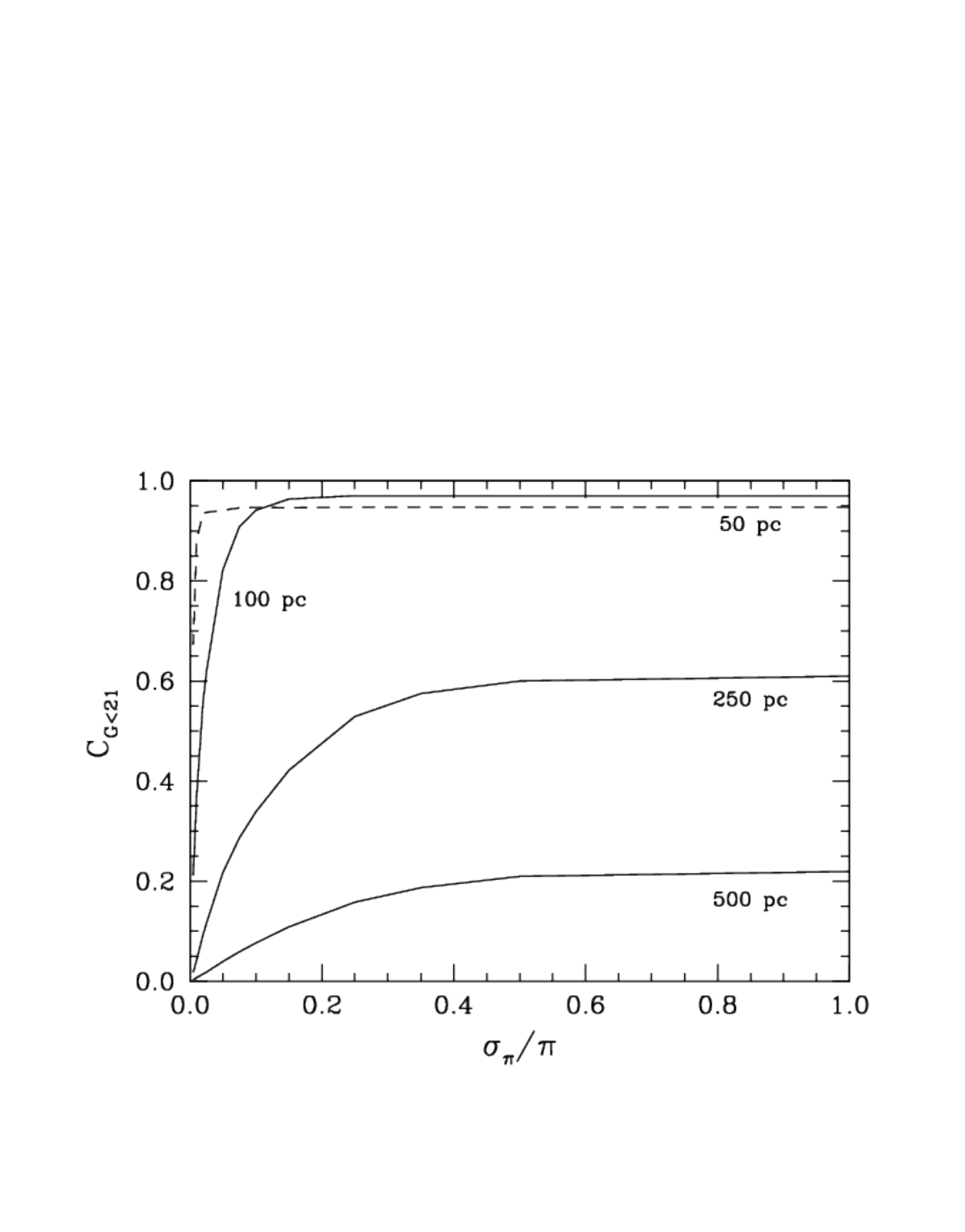}
\caption{Completeness of the white dwarf population sample accessible by {\it Gaia} as a function of the parallax relative error.}
\label{f-complet}
\end{figure}

First of all, we define the completeness for a given volume sample as the number of objects selected under certain selection criteria with respect to the total number of objects of that sample. Initially, we considered the white dwarf population accessible by {\it Gaia} (that is, assuming a magnitude limit of $G<21$) which presents a  parallax relative error, $\sigma_{\pi}/\pi$, below a certain value. The completeness obtained under these criteria is presented in Fig. \ref{f-complet} as a function of the parallax relative error and for different volumes.

For illustrative purposes, we considered different radii of our volume sample, i.e., 50, 100, 250, and 500\,pc. Our simulations show that for volumes larger than 250\,pc the completeness of the sample is below $\sim60\%$, and dramatically decreases to no more than $\sim22\%$ for sample sizes larger than 500\,pc. On the contrary, for samples up to 100\,pc the completeness can reach values as high as  $\sim97\%$.  As can be seen in Fig. \ref{f-complet}, the completeness for a 100\,pc sample remains practically constant for a parallax relative error greater than $\sim0.1$, while abruptly decreases for smaller values. Thus, we would expect that the vast majority of white dwarfs existing within 100\,pc will present a parallax error better than $10\%$.

\begin{figure*}
   \includegraphics[trim=4mm 25mm 16mm 20mm, clip,,width=1.6\columnwidth]{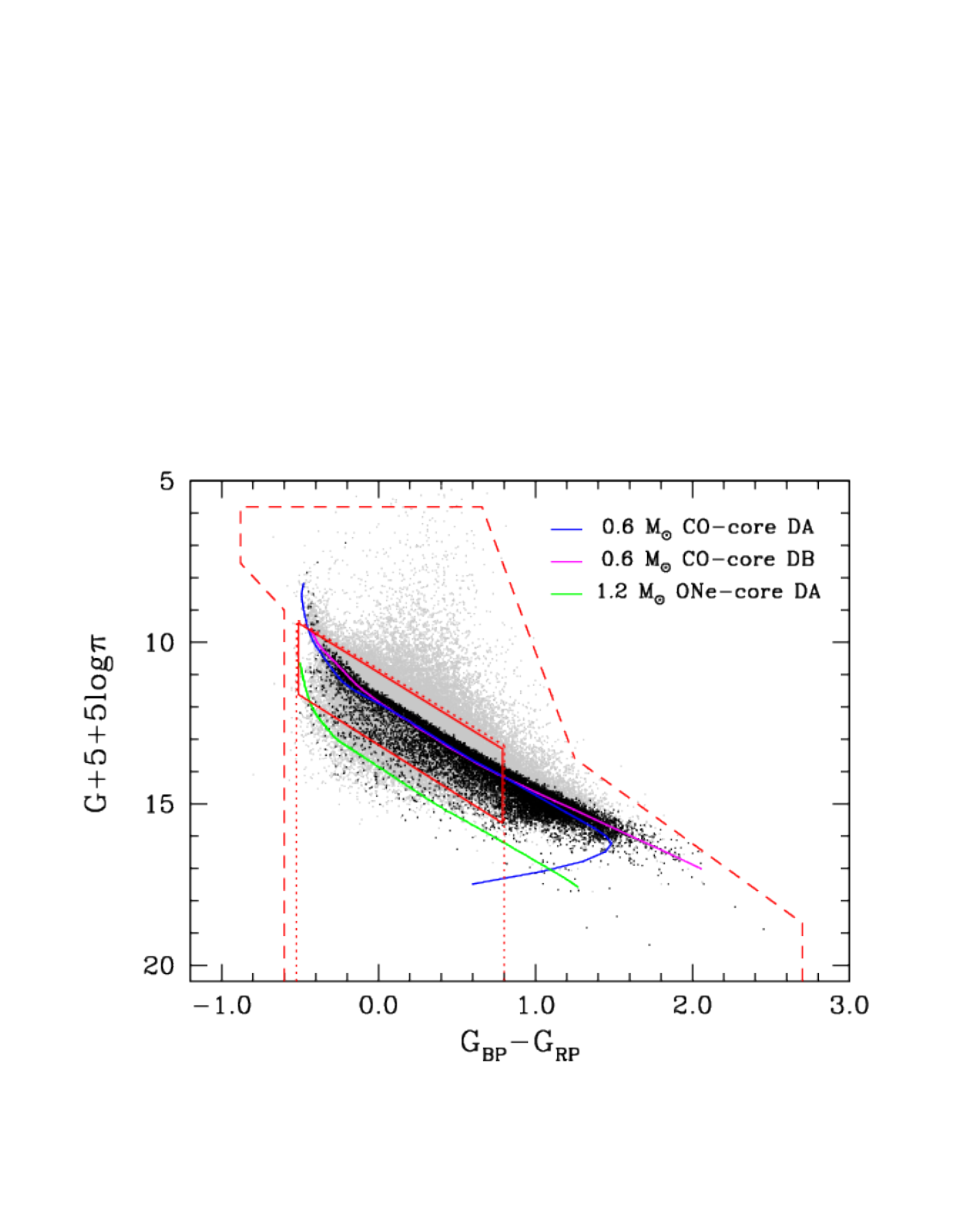}
\caption{{\it Gaia} G absolute magnitude - colour diagram for our synthetic white dwarf population accessible by {\it Gaia}. The whole white dwarf population is plotted by grey dots, while those within 100\,pc are plotted by black dots. Theoretical cooling tracks for a typical $0.6\,M_{\odot}$ (and 1.2$\,M_{\odot}$) white dwarf with different composition are shown with solid lines. The dashed lines define the region where all single white dwarfs accessible by {\it Gaia} were expected to be found. The dotted box is the same as the dashed one but for white dwarfs within 100\,pc and within our effective temperature range. And the red continuous box is the same as the dotted one but restricted for those white dwarfs with CO-cores (see text for details).}
\label{f-hrs}
\end{figure*}

It is also worth noting here that smaller samples are not necessarily more complete. This is also shown in our Fig. \ref{f-complet} by the completeness of the 50\,pc sample, marked for clarity reasons as a dashed line. The maximum completeness achieved by the 50\,pc sample, $95\%$, is slightly smaller than that of the 100\,pc sample. The reason that closer samples are not necessarily more complete is due to the Lutz-Kelker bias \citep{LutzKelker1973,Torres2007}. The parallax error associated to each object alters the observed spatial distribution of stars. Given that the volume of the sample increases as the square of the distance, the number of objects that enter a fixed volume shell does not have to be the same as the number of objects that leave that shell. In our case, we also needed to take into account that the density distribution of stars for the thin disc, thick disc, and halo population were different from each other and not constant. Consequently, it was not a priori obvious to deduce the effect of the parallax error on the final spatial distribution. However, our Monte Carlo simulations permitted us to clarify this issue and showed that, as previously mentioned, the completeness of the final sample does not increase for samples smaller than 100\,pc.

Consequently, and in view of our results of Fig. \ref{f-complet}, we adopted for our white dwarf catalogue a parallax error below $10\%$. This assumption guaranteed us a reasonable accuracy on distances and a $94\%$ completeness for the 100\,pc sample. That is to say, the magnitude limit of $G<21$ imposed by {\it Gaia} is in this case practically negligible, being this the reason why we can consider the 100\,pc sample as a volume-limited one, instead of a magnitude-limited one. Additionally, parallax errors below $10\%$ implied that the contribution to the absolute magnitude errors due to the distance uncertainty was not greater than 0.2 mag.

Secondly, we analysed the regions of the colour-magnitude diagram in which we expected to find white dwarfs. In Fig. \ref{f-hrs} we plotted our synthetic white dwarf population in the {\it Gaia} G absolute magnitude ($M_{G}=G+5+5\log \pi$) versus {\it Gaia} $G_{\rm BP}-G_{\rm RP}$ colour. The whole white dwarf population accessible by {\it Gaia} is plotted by grey dots, while those within 100\,pc are plotted by black dots. Also shown in Fig. \ref{f-hrs} for illustrative purposes are the theoretical cooling tracks for a typical $0.6\,M_{\odot}$ CO-core hydrogen-rich atmosphere \citep{Renedo2010} -- blue line --, a $0.6\,M_{\odot}$ CO-core hydrogen-deficient atmosphere \citep{Althaus2005} -- magenta line --, and a $1.2\,M_{\odot}$ ONe-core white dwarf \citep{Althaus2007} -- green line. 

A first feature that can be highlighted, in view of the results of Fig. \ref{f-hrs}, is the extend to brighter magnitudes of the white dwarf population when compared to those objects which lie within 100\,pc. This fact, by no means, implies that distant objects are intrinsically brighter than those closer than 100\,pc. Being {\it Gaia} a magnitude-limited sample, it is natural that the selection is biased towards intrinsically brighter stars as the observing distance increases. Additionally, smaller parallaxes (larger distances) imply larger relative errors and less precise absolute magnitudes. For instance, an object with a 10\% error in its parallax estimate would present a magnitude spread (only due to the trigonometric error) of $\Delta M\approx 0.2\,$mag, $\approx 0.5\,$mag for a 25\% parallax error, and more than 1\,mag for a 50\% parallax error. Consequently, what we observe  in Fig. \ref{f-hrs}, is the effect of the large dispersion in absolute magnitudes induced by larger parallax relative errors for those objects far beyond 100\,pc.

From Fig.\,\ref{f-hrs}, we established certain regions where we expected to find the white dwarf population. First, we delimited a wide region where all single white dwarfs accessible by {\it Gaia} can be found. This big box is marked by a red dashed line in Fig. \ref{f-hrs}. Secondly, we defined a region where we expected to find white dwarfs within 100\,pc ($M_{G}\,>\,2.95\times(G_{\rm BP}-G_{\rm RP})+10.83$). This sample, we recall, is the most complete (see Fig.\,\ref{f-complet}). However, our adopted white dwarf model atmosphere spectra only cover the range of effective temperatures from 5,000\,K to 80,000\,K (see Section \ref{Koestermodels}). We thus defined a colour region to select 100\,pc white dwarfs for such effective temperatures. It is important to mention though that, as we demonstrate in Section \ref{query}, the expected contamination in our catalogue for effective temperatures lower than 6,000\,K is very large. Consequently, we adopted a conservative lower effective temperature limit of 6,000\,K, rather than 5,000\,K, and an upper limit of 80,000\,K. This added the colour condition $-0.52<(G_{\rm BP}-G_{\rm RP})<0.80$ to our sample. The magnitude-colour region thus obtained is marked as red dotted lines in Fig. \ref{f-hrs}. 

We narrowed our selection region further taking into account the mass of the white dwarfs. This is because more massive white dwarfs have smaller radii and consequently, for a fixed effective temperature, they are fainter \citep[e.g.][]{Provencal1998,Althaus2010a,Parsons2017}. That is clear in Fig. \ref{f-hrs} by the locus of the ONe-core white dwarf cooling track when compared to the brighter cooling track for a CO-core white dwarf. Thus, we delimited a small box (continuous red line in Fig. \ref{f-hrs}) only formed by CO-core white dwarfs by separating those ONe-core white dwarfs, i.e., those with $M_{G}>\,3.02\times(G_{\rm BP}-G_{\rm RP})+13.18$. This delimiting line, which approximately follows the cooling sequence of a $\approx 1.04\,M_{\odot}$ white dwarf, does not pretend to be a strict classification, and should be only understood for guiding purposes. It is worth saying here that the CO-core white dwarf population finally obtained in the small box is formed by both DA and DB stars. In fact, the region occupied by these two subpopulations within this box is indistinguishable. The blue turn, that occurs on hydrogen-rich atmospheres due to the collision-induced absorption on molecular hydrogen, occurs at lower temperatures, $\lappr\,5,000\,$K. That region, which eventually may permit us to distinguish between DA and non-DA white dwarfs is, as clear seen in the Fig. \ref{f-hrs}, beyond our lower temperature limit. Consequently, we expect some contamination of non-DA stars in the final sample, which would affect the effective temperature distribution, inasmuch as our model atmosphere spectra are specific for DA white dwarfs (see Section \ref{Koestermodels}).

\begin{table}
\caption{Completeness estimate of the white dwarf population for different volume samples after cumulatively applying our three selection cuts. See text for details. }
\begin{center}
\begin{tabular}{ccccc}  
\hline \hline Selection cut & 50 pc & 100\,pc & 250 pc &  500 pc \\
\hline 
$G<21$ & 0.95 & 0.97 & 0.61 & 0.22 \\
$\sigma_{\pi}/\pi$<0.1 & 0.95 & 0.94 & 0.34 & 0.08 \\
6,000\,K\,$<T_{\rm eff}<$\,80,000\,K  & 0.43 & 0.44  & 0.27 & 0.04 \\
\hline \hline
\end{tabular}
\end{center}
\label{t:table1}
\end{table}

Finally, in Table \ref{t:table1} we summarise the estimated completeness of the white dwarf population as a function of the different selection cuts employed and for different volume samples. In particular, for our reference sample of 100\,pc, the initial completeness for the population accessible by {\it Gaia} was a $97\%$. The selection of those stars with a parallax error smaller than a $10\%$, slightly reduced the completeness to a $94\%$. Lastly, the effective temperature cut (i.e. colour cut) reduced the completeness of the final sample to $44\%$. It is also worth noting that smaller samples, such as the 50\,pc sample obtain similar values for the final completeness. However, larger volumes, such as the 250 or 500\,pc samples presented in Table \ref{t:table1}, exhibit significantly lower completenesses, $27\%$ and $4\%$, respectively. Definitely, our reference sample of 100\,pc analysed so far presents the highest completeness for the larger distance considered.

\begin{figure*}
   \includegraphics[trim=16mm 30mm 16mm 40mm, clip,,width=1.6\columnwidth]{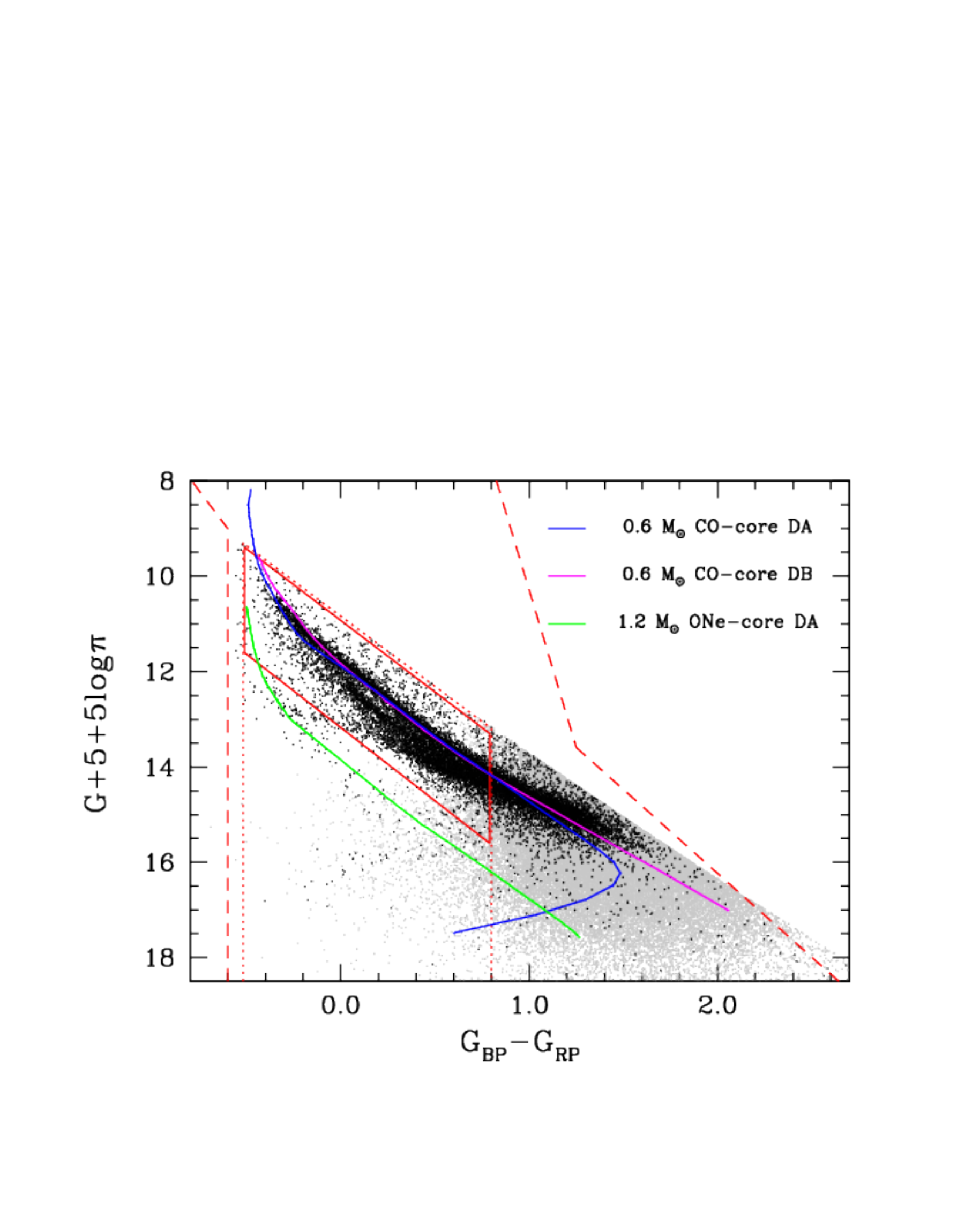}
\caption{{\it Gaia} G absolute magnitude - colour diagram for the {\it Gaia} white dwarf population within 100\,pc. Those objects that do not fulfill our colour excess criterion are marked as grey dots. Theoretical cooling tracks are the same as in Fig.\ref{f-hrs}.}
\label{f-hr-excess}
\end{figure*}
\tR{}

\subsection{Query to the {\it Gaia}-DR2 catalogue}
\label{query}

According with the discussion in Sect.\ref{s-criteria} and the recommendations in \cite{Evans18} and \cite{Lindegren18}, we searched for white dwarf candidates in the {\it Gaia} DR2 catalogue\footnote{http://gea.esac.esa.int/archive/} using the following criteria:

\begin{itemize}
\item $\pi>0$\footnote{Excluding negative parallaxes does not affect the main results of our paper since we focus our analysis on the 100\,pc white dwarf sample.} and $\pi/\sigma_{\pi}>10$
\item $F_{\rm BP}/\sigma_{F_{\rm BP}}>10$ and $F_{\rm RP}/\sigma_{F_{\rm RP}}>10$
\item ${\it phot\_bp\_rp\_excess\_factor}<1.3+0.06\times(G_{\rm BP}-G_{\rm RP})^{2}$ 
\item $-\,0.52<(G_{\rm BP}-G_{\rm RP})<0.80$
\item $2.95\times(G_{\rm BP}-G_{\rm RP})+10.83<M_{G}<3.02\times(G_{\rm BP}-G_{\rm RP})+13.18$ for CO-core white dwarfs
\item $M_{G}>3.02\times(G_{\rm BP}-G_{\rm RP})+13.18$ for ONe-core white dwarfs
\end{itemize}

As we have discussed above, the 10\% uncertainty in the parallax corresponds to an uncertainty of $\sim$0.2 mag in the absolute G magnitude. And, the 10\% of uncertainty of the {\it Gaia} fluxes to an uncertainty of $\sim$0.14 mag in the {\it Gaia} $G_{\rm BP}-G_{\rm RP}$ colour. Since the G band measurements have lower errors, we did not impose any cut in this band. 

The cut in the {\it phot\_bp\_rp\_excess\_factor} prevented against photometric errors in the BP and RP bands, which are especially important for faint sources in crowded areas. In order to evaluate the effect of this criterion, we performed a check by selecting {\it Gaia} objects within 100\,pc regardless of the colour limits. The magnitude-colour diagram thus obtained is shown in Fig. \ref{f-hr-excess}. Those objects which fulfill our photometric excess factor criterion are marked as black dots, while those which exceed it are displayed by grey dots. The magnitude-colour diagram obtained for the objects under our photometric criterion greatly resembles the one predicted by our population synthesis model presented in Fig. \ref{f-hrs}. On the other hand, the objects that do not fulfill our colour excess criterion are mainly concentrated at $G_{\rm BP}-G_{\rm RP}>0.8$ and fainter magnitudes. In view of Fig. \ref{f-hr-excess}, it is clear that this region is strongly contaminated and deserves a thorough analysis, being this another reason why we did not include it in this first presentation of our catalogue. 

Our colour excess criterion excluded a significant fraction of objects in the region where ONe-core white dwarfs were expected. Likewise, a few objects inside the CO-core box were erased. Nevertheless, this effect is only partially important for cool white dwarfs ($6,000$ K\,$\lappr\,T_{\rm eff}\,\lappr\,8,000$ K), being it much more intense for very cool objects ($T_{\rm eff}\,\lappr\,6,000$ K). In fact, for temperatures below 5,000 K the colour excess dominates the sample. For these reasons we preferred to avoid this region, despite the fact that atmospheric models with temperatures down to 5,000 K are available in the adopted grid (see Sect. \ref{Koestermodels}). 

Consequently, we concluded that our colour limit of $(G_{\rm BP}-G_{\rm RP})<0.80$, equivalent to a effective temperature cut of $\approx\,6,000$ K, together with the adopted photometric criteria, permitted us to obtain a catalogue of white dwarf candidates with low contamination and discarding the minimum number of objects.

\section{The {\it Gaia} White Dwarf catalogue}
\label{results}

We proceeded to extract our white dwarf candidates from the {\it Gaia} DR2 data applying the searching criteria exposed in Sect.\,\ref{query}. Our query returned 73,221 {\it Gaia} DR2 counterparts, 72,178 CO-core and 1,043 ONe-core white dwarf candidates. They all form our catalogue of {\it Gaia} white dwarfs. All the relevant information of these sources can be gathered from {\em The SVO archive of {\it Gaia} white dwarfs} at the Spanish Virtual Observatory portal\footnote{http://svo2.cab.inta-csic.es/vocats/v2/wdw} (see Appendix~\ref{append}).

Of the total sample, 8,343 CO-core and 212 ONe-core were within 100\,pc. Once corrected from completeness (see Table \ref{t:table1}), we derived from these values a white dwarf spatial density of $4.9\pm0.4\times10^{-3}\,{\rm pc^{-3}}$ for the 100\,pc sample. This value is in perfect agreement with the spatial density of $4.8\pm0.5\times10^{-3}\,{\rm pc^{-3}}$ derived from  the 20 pc volume-limited ($86\%$ complete) sample or the same value obtained for the extended version up to 25 pc ($68$\% complete)  of \cite{Holberg2008,Holberg2016}. This fact reinforces our previous assumption that our catalogue up to 100\,pc forms a complete volume-limited sample in the range of colours considered.

\begin{figure}
   \includegraphics[width=\linewidth]{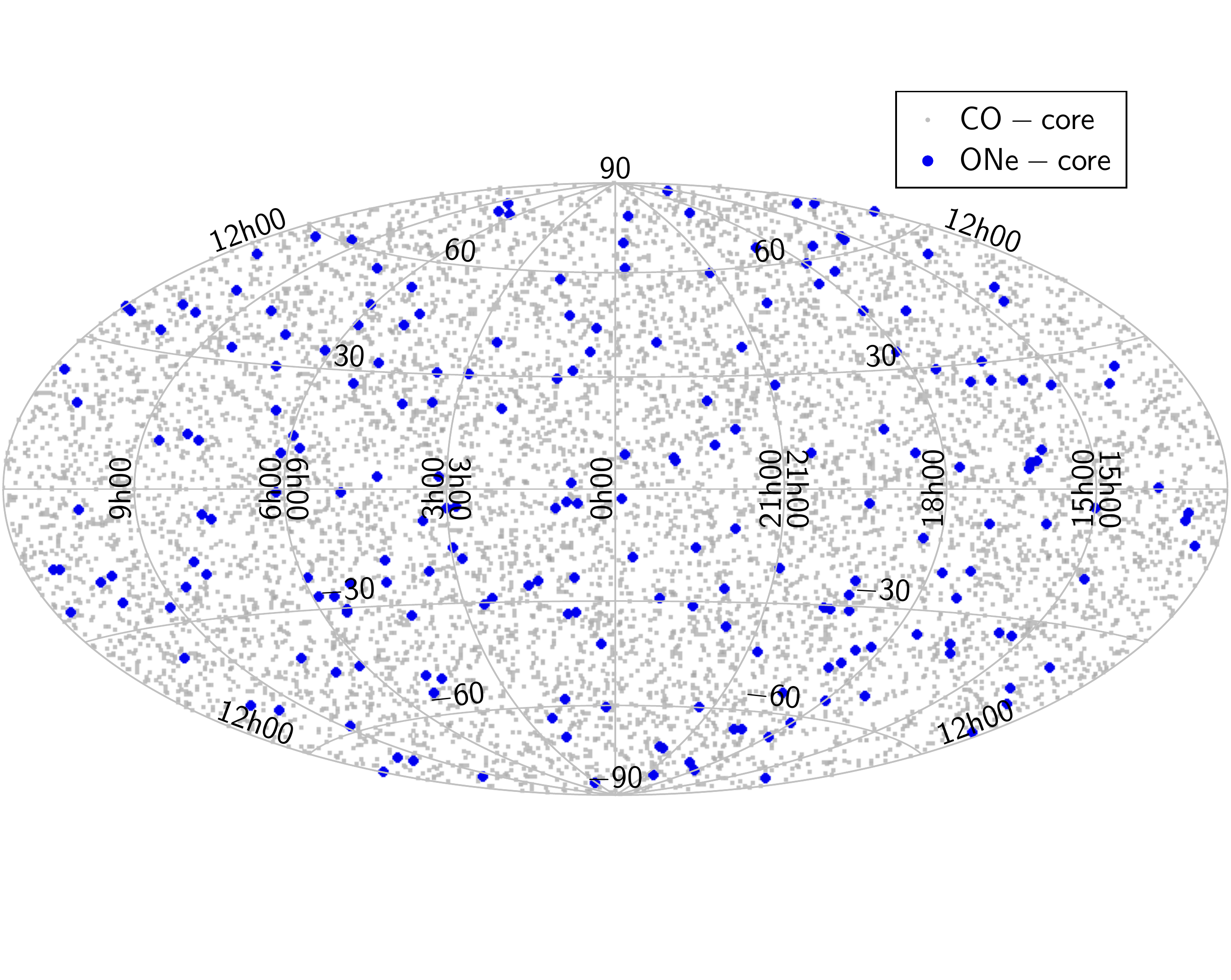}\par 
\caption{Sky distribution in equatorial coordinates of the 100\,pc sample candidate white dwarfs.}
\label{f-sky100}
\end{figure}

\begin{table*}
\caption{Available SIMBAD classification for 15,930 of our selected {\it Gaia} white dwarfs. We provide the number of members in each class for both the full catalogue and the 100\,pc sub-sample. Classifications in agreement with the white dwarf nature of our selected candidates are shown in italics. The estimated contamination was $\lappr$1\% for both the whole and the 100\,pc samples.}
\begin{center}
\begin{tabular}{crrcrr}
\hline
\hline
Simbad class. & N  & 100\,pc & Simbad class. & N & 100\,pc \\
\hline
\emph{WD}           & 10,356 & 2,218 & sdB          &  42 &  19 \\
\emph{pulsating WD} &     53 &    24 & Radio        &   1 &     \\
\emph{Blue object}  &  3,409 &   219 & Transient    &   1 &     \\
\emph{UV source}    &    866 &    66 & Galaxy       &  30 &   1 \\
\emph{post AGB}     &      1 &       & Quasar       &  38 &   1 \\
  AMHer             &      2 &       & AGN          &   1 &     \\
  CV                &     11 &       & BLLac        &  18 &   4 \\
  Dwarf Nova        &      9 &     2 & Variable star &   4 &   1 \\
  Nova              &      6 &     2 & Algol          &   1 &     \\
  Novalike          &      1 &     1 & High proper motion & 815 & 596 \\
  Emission line     &      1 &     1 & Long Period Variable      &   1 &   1 \\
  Emission object   &      4 &     1 & In Binary/multiple system &   33  &   16 \\
  Eruptive          &      1 &     1 & IR   &  1  &    \\
\hline
\end{tabular}
\end{center}
\label{t:class}
\end{table*} 

Figure \ref{f-sky100} shows the sky distribution in equatorial coordinates of the 100\,pc sample for our CO-core (grey dots) and ONe-core (blue dots) white dwarf candidates. The distribution of these sources reveals the full sky coverage of {\it Gaia} performance in a nearly homogeneous distribution, as it is expected for a volume-limited local sample of stars. Additionally, the fact that the sky distribution does not show concentration to high density regions is an indication that the selection criteria were effective removing contamination.

\subsection{Empirical completeness}

The numerical simulations performed in Sect.\ref{s-criteria} allowed us to estimate a completeness of 94\% for the synthetic 100\,pc white dwarf sample accessible by {\it Gaia}. In this section we derive an empirical estimate of the completeness of the {\it Gaia} selected sample for the same distance volume. To that end we made use of the largest current spectroscopic catalogue of white dwarfs from the SDSS with available distance estimates \citep{Anguiano17}, for which the range of effective temperatures is similar to our range. From this list of $\sim$21,000 objects, we considered 388 CO-core white dwarfs with distances below 100\,pc and a distance relative error of less than 10\%. Our {\it Gaia} catalogue contained all but 24 SDSS white dwarfs. In other words, $\sim$94\% of the SDSS CO-core white dwarf sample has been identified by {\it Gaia} and have a parallax relative error below 0.1, in perfect agreement with our expectations. Visual inspection of the SDSS spectra and images of these 24 sources did not reveal any obvious reason why these objects were not recovered by {\it Gaia}.

\subsection{Contamination}  

In order to evaluate possible contamination of other sources in our white dwarf catalogue, we searched for available SIMBAD counterparts within 3\arcsec\ of our selected sources. 16,452 counterpart were found, of them 3301 are within 100\,pc. SIMBAD provided classification for all of them but 522, of which 129 are within 100\,pc. The results are shown in Table\,\ref{t:class}. If we assume SIMBAD to always provide correct classifications, and by looking at the Table, one can clearly see that most ($\sim$92\%, $\sim$80\% within 100\,pc) of the classified objects are indeed white dwarfs (or at least they have a classification in agreement with the objects being white dwarfs, i.e. blue objects or UV sources). Approximately, another 7\% (19\% within 100\,pc) of sources had a classification not discriminatory of white dwarf nature (i.e. high proper motion). The highest sources of contamination seem to be hot subdwarfs ($\sim$0.3\%, $\sim$0.6\% within 100\,pc), cataclysmic variables ($\sim$0.17\%, $\sim$0.16\% within 100\,pc), and galaxies/quasars ($\sim$0.6\%, $\sim$0.2\% within 100\,pc). However, given that hot white dwarfs have similar colours to quasars, the classification provided by SIMBAD is likely to be incorrect in these cases and these objects are most probably white dwarfs. Thus, the expected contamination in our white dwarf catalogue is very low ($\lappr$1\%).


\subsection{The {\it Gaia} white dwarf Hertzsprung-Russell diagram}
\label{GaiaHR}

\begin{figure*}
  \includegraphics[trim=1mm 10mm 6mm 20mm, clip,,width=1.8\columnwidth]{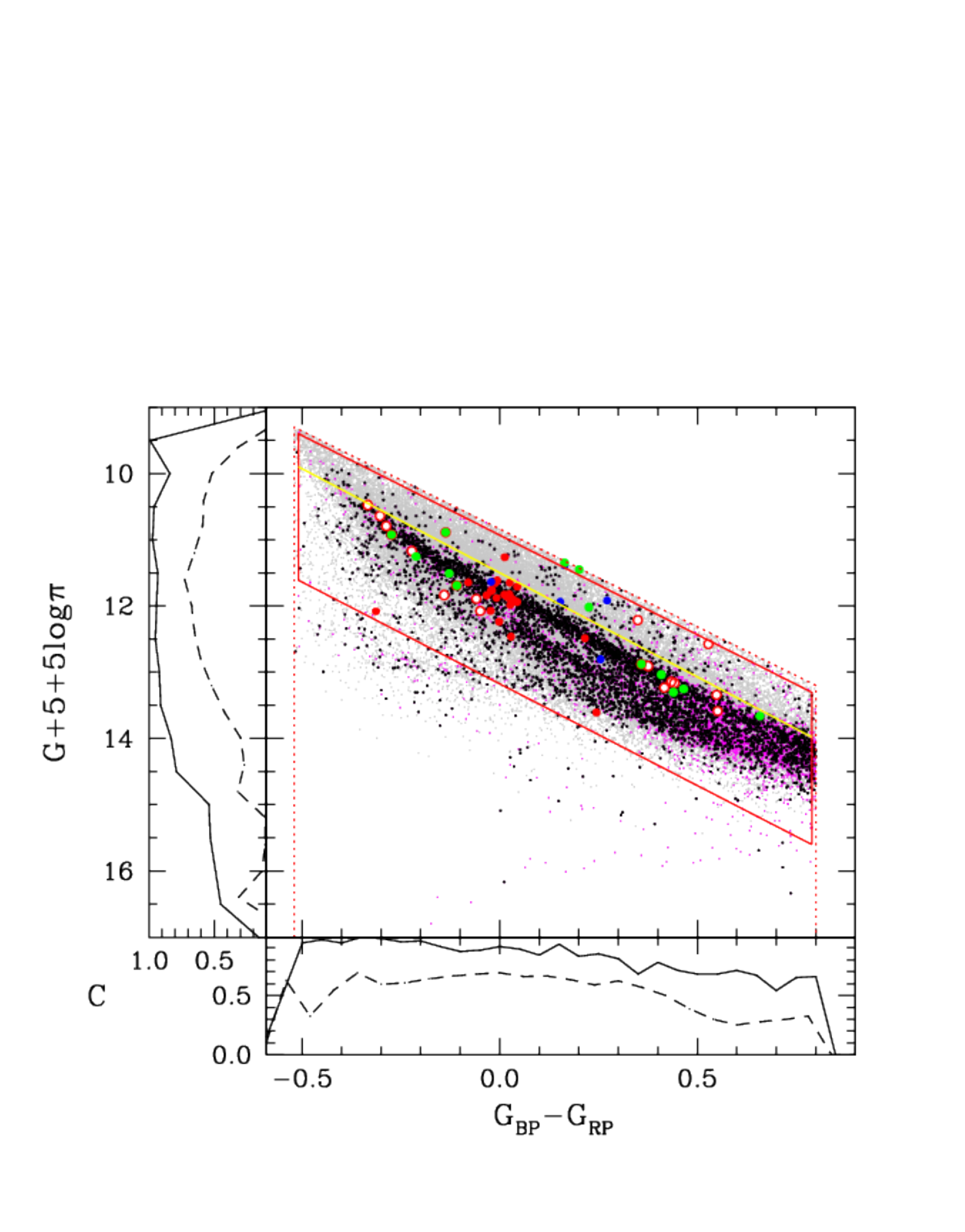}
\caption{{\it Gaia} G absolute magnitude - colour diagram for the white dwarfs identified in our catalogue (central panel). Objects beyond 100\,pc are marked as grey dots, those which belong to our 100\,pc sample and present good estimates of its physical parameters are plotted as black dots, while those within 100\,pc and poor estimates are represented by magenta dots. For visual purposes, the yellow continuous line marks the region above which unresolved double-degenerate binaries or He-core white dwarfs are expected to be located for the 100\,pc sample. Also represented by circles are some of the objects classified by SIMBAD: pulsating white dwarfs (red), Novae (blue), sdB (white) and binary systems (green). The completeness of the 100\,pc sample as function of the $G_{\rm BP}-G_{\rm RP}$ colour (bottom panel) and the G absolute magnitude (left panel) is shown by a continuous line. The completeness for those objects with good VOSA fits (see Section\,\ref{vosa}) is marked with a dashed line.}
\label{f-hrcat}
\end{figure*}

The {\it Gaia} Hertzsprung-Russell diagram for our catalogue is presented in the central panel of Fig. \ref{f-hrcat}. White dwarf candidates beyond 100\,pc are marked as grey dots, those that belong to our 100\,pc sample and present good physical parameter estimates (see Section \ref{vosa}) are plotted as black dots, and those within 100\,pc and poorly determined parameters are represented by magenta dots. Also included are some of the objects with available SIMBAD classification: pulsating white dwarfs (red solid circles), Novae (blue solid circles), sdB (open white circles) and binary systems (green solid circles). The completeness of the 100\,pc sample was derived as a function of the $G_{\rm BP}-G_{\rm RP}$ colour and the G absolute magnitude, as shown by a continuous line in the bottom and left panels of Fig. \ref{f-hrcat}. The completeness achieved within the range of colours considered was quite close to be 100\%, with the contamination being negligible as stated in the previous sections. The average completeness of our 100\,pc sample was nearly a $93\,\%$, in agreement with our estimates of Section\,\ref{s-criteria}. Additionally, we plotted in Fig.\,\ref{f-hrcat}, by means of a dashed line, the corresponding completeness for those objects with good stellar parameter determinations derived from VOSA (see Section\,\ref{vosa}). In this case, the completeness of our sample fell to a $63\,\%$, being specially reduced for red colours $G_{\rm BP}-G_{\rm RP}>0.5$, or equivalently for low effective temperature objects, $T_{\rm eff}$\,$\lappr$\,8,000 K.

A general glance at the central panel of Fig.\ref{f-hrcat} reveals a good agreement with the expected distribution for our synthetic model of the Galactic white dwarf population shown in Fig. \ref{f-hrs}. In particular, this match indicates that our treatment of the photometric and astrometric errors was in agreement with {\it Gaia} observations. However, some discrepancies with current models require  a further analysis.

\begin{figure*}
  \includegraphics[trim=25mm 1mm 27mm 1mm, clip,,width=1.\columnwidth]{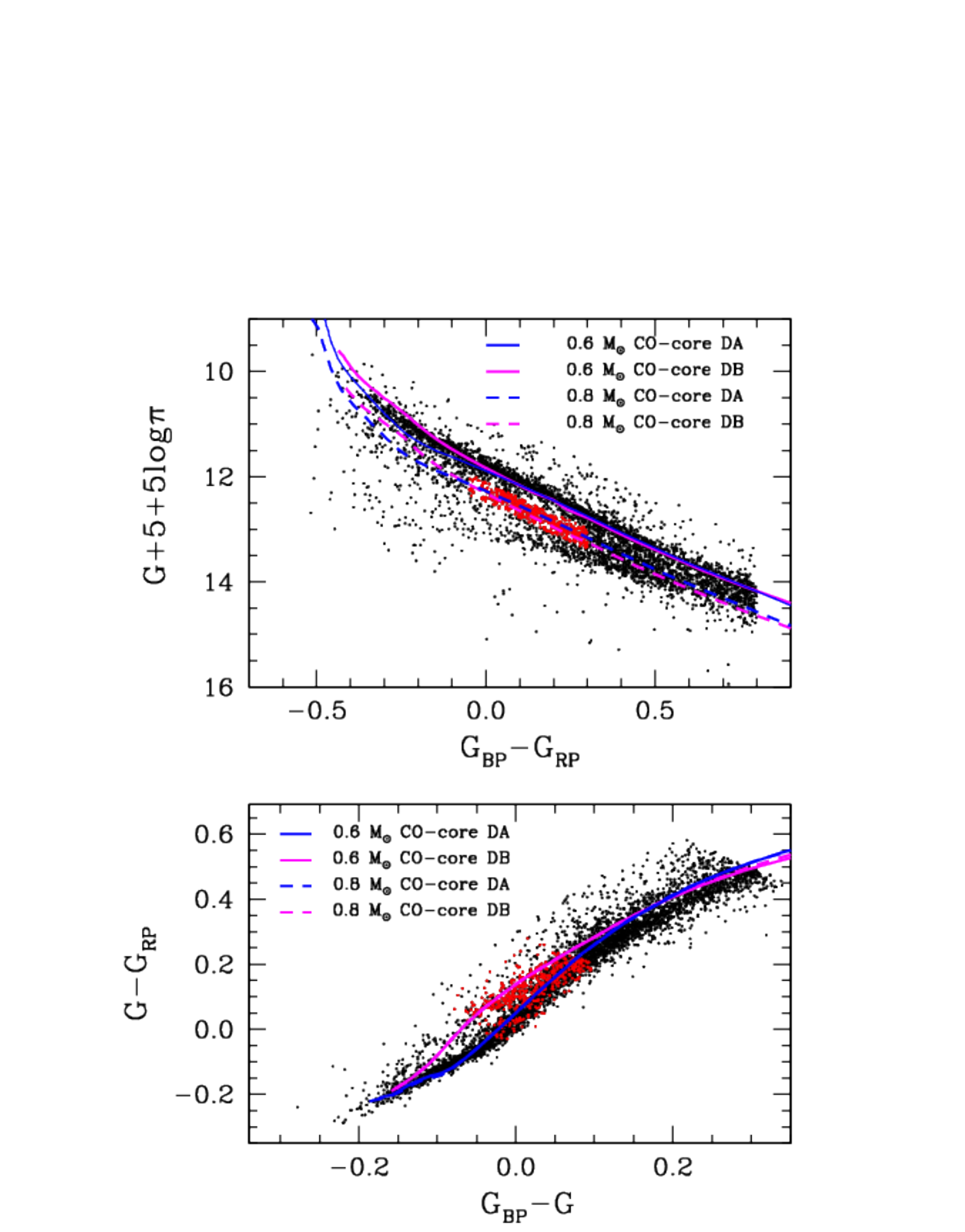}
    \includegraphics[trim=25mm 1mm 27mm 1mm, clip,,width=1.\columnwidth]{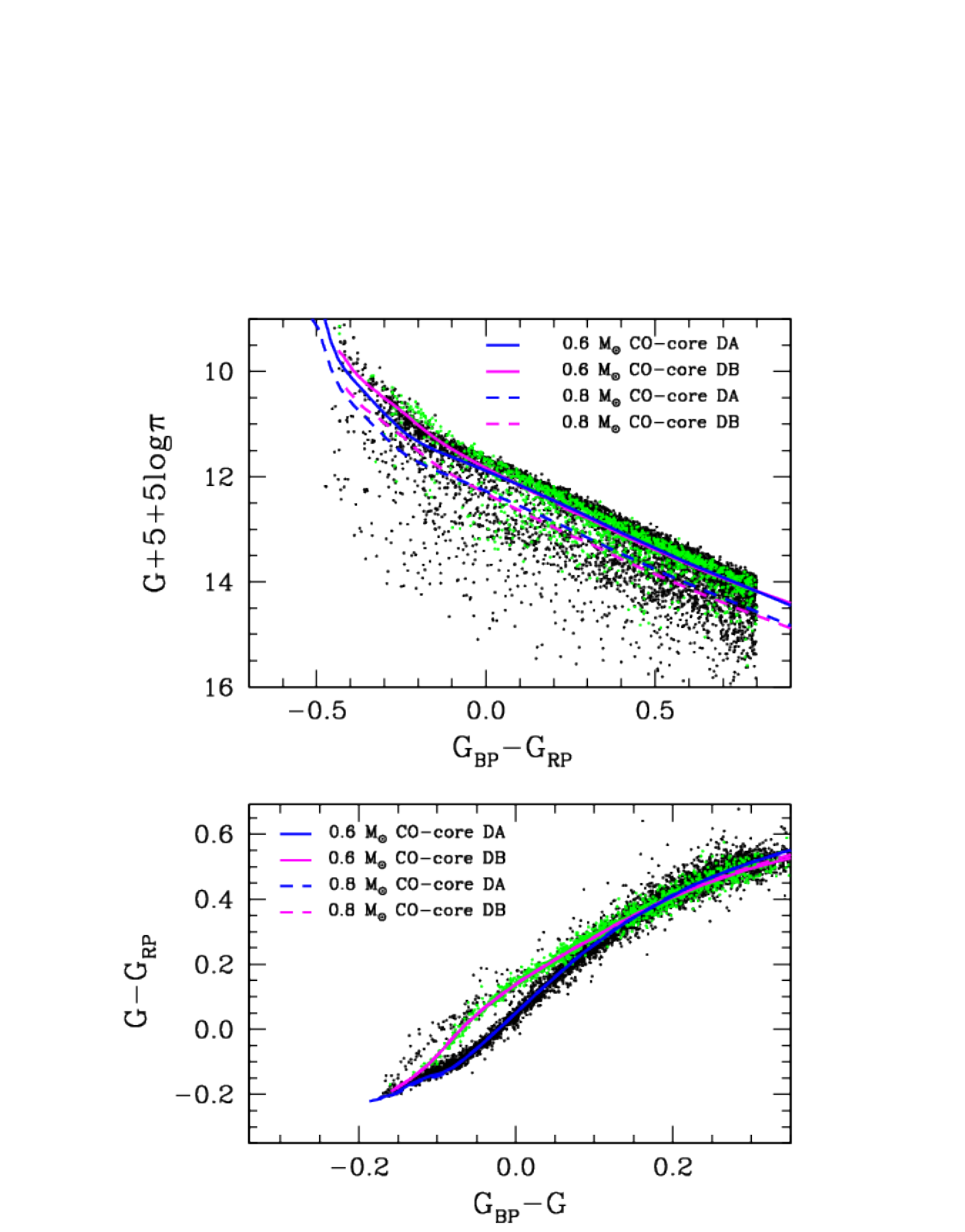}
\caption{{\it Gaia} G absolute magnitude - colour diagram (top-left) and colour-colour diagram (bottom-left) for the white dwarfs identified in our 100\,pc catalogue sample. Objects belonging to the low bifurcation region are marked as red dots for their identification. Theoretical cooling tracks for a $0.6\,M_{\odot}$ both DA and DB white dwarfs are shown with solid lines, and for $0.8\,M_{\odot}$ with dashed lines. The results obtained from our population synthesis models (right panels) are not able to reproduce the observed bifurcation in the magnitude-colour diagram, but they reproduce instead the observed colour-colour diagram. Consequently, the population of the lower branch of the bifurcation  can not be completely attributed to helium-rich atmosphere white dwarfs (green dots). See text for details.}
\label{f-hrcc}
\end{figure*}

First of all, one apparent trend presented in the {\it Gaia} Hertzsprung-Russell diagram of Fig.\ref{f-hrcat} for the 100\,pc population is the clear bifurcation around $0.0\,\lappr\,G_{\rm BP}-G_{\rm RP}\,\lappr\,0.3$. This bifurcation is not expected in current white dwarf Galactic models, as shown in our synthetic magnitude-colour diagram of Fig. \ref{f-hrs}, and can not be attributed to any observational bias nor photometric or astrometric error. This fact, already discussed by the {\it Gaia}-collaboration \citep{Babusiaux2018}, deserves a more thorough analysis. In order to do this, we plotted in Fig.\,\ref{f-hrcc} the HR-diagram (top-left panel) and the colour-colour diagram (bottom-left panel) for the 100\,pc sample {\it Gaia} identified white dwarfs. In the right panels of the same Figure we show the corresponding diagrams for our synthetic population. For comparative reasons, we draw the theoretical cooling tracks for a typical $0.6\,$M$_{\odot}$ CO-core white dwarf with a hydrogen-rich, DA, atmosphere (blue continuous line) and with a hydrogen-deficient, DB, atmosphere (magenta continuous line). Also plotted are the corresponding tracks for a $0.8\,$M$_{\odot}$ DA (blue dashed line) and DB (magenta dashed line) white dwarf. For a better identification, DB white dwarfs are marked as green dots in our simulated diagrams. 

A closer look to the HR-diagrams (top panels of Fig.\,\ref{f-hrcc}) reveals that the theoretical cooling tracks for typical 0.6\,$M_{\odot}$ white dwarfs for both DA and DB models are superimposed in the bifurcation region. The predicted theoretical bifurcation between white dwarf atmospheric models occurs at hotter temperatures, or equivalently bluer colours, $G_{\rm BP}-G_{\rm RP}\sim -0.2$. Furthermore, the small discrepancies, if any, observed between the average mass of DA and DB white dwarfs (see \citealt{Bergeron2011,Giammichele2012,Kepler16} and reference therein) are by no means able to reproduce the observed bifurcation. The lower branch of this bifurcation implies more massive objects. However, massive cooling tracks for DA and DB objects are also superimposed. 

In order to break this degeneracy and to disentangle whether or not the observed bifurcation is produced by DA or DB white dwarfs, we required to analyse the distribution in the colour-colour diagram (bottom panels  of Fig.\,\ref{f-hrcc}).  In this diagram the different types of atmospheric models are split into distinct cooling curves. For a better identification of the objects belonging to the lower branch of the bifurcation, those between $-0.05\,\lappr\,G_{\rm BP}-G_{\rm RP}\,\lappr\,0.3$ and 
$2.95\times(G_{\rm BP}-G_{\rm RP})+12.1<M_{G}<3.02\times(G_{\rm BP}-G_{\rm RP})+12.5$, we marked them as red dots. The location of these objects within the colour-colour diagram (bottom left panel of Fig.\,\ref{f-hrcc}) clearly reveals that not all of them are helium-rich atmosphere white dwarfs, as some initial studies claim \citep{Babusiaux2018}. In fact, our estimates give a rough $30-40\,\%$ of DB white dwarfs among the sources belonging to the lower branch of the bifurcation region. Hence, this characteristic feature represents an important discovery of the 100\,pc white dwarf population extracted from the {\it Gaia} catalogue. 

A second minor discrepancy between the observed and simulated HR diagrams arises for those objects of the {\it Gaia} 100\,pc sample located between the upper limit of our CO-core small box ($M_{G}=2.95\times(G_{\rm BP}-G_{\rm RP})+10.83$) in Fig. \ref{f-hrcat} and the main white dwarf sequence, which are not reproduced by our simulations (Fig. \ref{f-hrs}). The specific location of these objects within the HR-diagram can be attributed to a double origin. One possibility is that these objects are single He-core white dwarfs. Although the existence of He-core white dwarfs is generally associated to binary evolution, there remains a small fraction of $\sim\,15\%$ with unclear origin (see \citealt{Rebassa11} and references therein). A second possibility consist in unresolved double-degenerate binaries. The sum of the individual fluxes of the two white dwarfs in the binaries results in brighter (unresolved) objects. In any case, further studies beyond the scope of this work are necessary to ascertain the true nature of these objects.

Finally, from Fig.\,\ref{f-hrcat} we can check that those objects with bad photometric parameters (magenta dots) are more numerous for redder colours. This effect joins the one previously commented in Section\,\ref{query}, i.e. the excess in colour criterion is also more significant for redder objects. Both effects contribute to eliminate more white dwarfs from our 100\,pc sample as we move closer to our low temperature limit. This is clearly reflected in the bottom panel of Fig.\,\ref{f-hrcat}  by the drop of completeness for redder colours.


\section{The VOSA SED fitting}
\label{vosa}

VOSA\footnote{http://svo2.cab.inta-csic.es/theory/vosa} \citep{Bayo08} is a Virtual Observatory tool designed to determine physical parameters (such as effective temperature and luminosity) of thousands of objects at once. VOSA compares observed photometry, gathered from a significant number of VO  complaint catalogues, to different collections of theoretical models. 

For this work, we queried the GALEX \citep{Bianchi00}, {\it Gaia} DR2 \citep{Brown18}, APASS DR9 \citep{Evans02}, SDSS DR12 \citep{Alam15}, Pan-STARRS1 \citep{Chambers16}, ALHAMBRA \citep{Moles08}, the Dark energy Survey (DES) \citep{DESC16}, 2MASS \citep{Skrutskie06}, VISTA \citep{Cross12}, and UKIDSS \citep{Hewett06} photometric catalogues. For each catalogue, different search radius were tested in order to obtain the maximum number of counterparts avoiding contamination. Thus, a search radius between 2\arcsec\ and 3\arcsec was selected, depending on the astrometric accuracy of the queried catalogue. We used the {\it Gaia} coordinates corrected for proper motion and transformed to J2000 provided by the CDS. 

This way we built the observational SEDs from the UV to the NIR wavelength range, where most of our objects radiate most of their light. The observational SEDs were then compared to a grid of synthetic spectra specifically developed for white dwarfs (see Section \ref{Koestermodels}). We did not account for interstellar extinction since it was unknown for most of our sample. Nevertheless, as the sources in our sample are located at short distances the effect of the extinction was expected to be small. Two physical parameters were obtained from the SED fitting: effective temperature and luminosity. This last was obtained from the bolometric flux and the distance obtained from the {\it Gaia} parallax. To calculate the bolometric flux, VOSA integrates the flux using the observational photometric points, and the theoretical model that best fits at the region of the SEDs not covered by the observations. This allows to obtain a more accurate bolometric flux, and so luminosity, than from bolometric corrections. In addition, the effective temperature was obtained from the SED fitting instead than from the {\it Gaia} colour, which minimises the impact of photometric errors at the {\it Gaia} bands. 

\subsection{Synthetic white dwarf atmosphere emission spectra}
\label{Koestermodels}
 
As said above, in order to derive the physical parameter of the studied objects, VOSA requires a set of synthetic spectra to compare to the observational SEDs. To study our sample of candidate white dwarfs, we adopted the updated DA white dwarf model spectra of \citet{Koester10}, for which the parameterization of convection follows the mixing length formalism $ML2/T=0.7$. The grid contains synthetic spectra of effective temperatures ranging from 5,000\,K to 20,000\,K in steps of 250\,K, from 20,000\,K to 30,000\,K in steps of 1,000\,K, from 30,000\,K to 40,000\,K in step of 2,000\,K, from 40,000\,K to 50,000\,K in steps of 5,000\,K, and from 50,000\,K to 80,000\,K in steps of 10,000\,K, and surface gravities ranging between 6.5 and 9.5\,dex in steps of 0.25\,dex for each effective temperature.

\begin{figure}
   \includegraphics[width=\columnwidth]{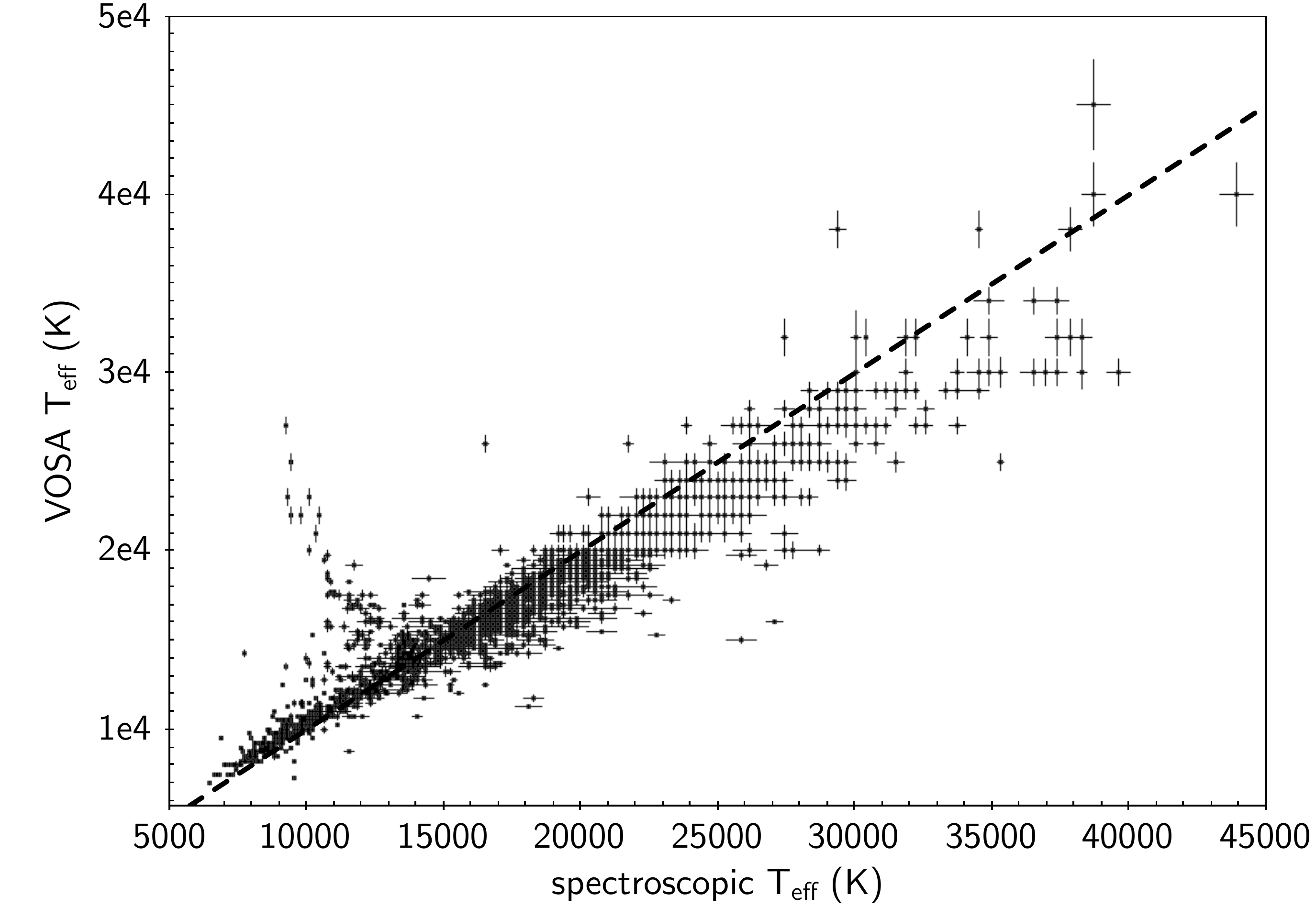}
\caption{Comparison between the spectroscopic SDSS and the photometric VOSA effective temperatures.}
\label{f-sdssvosa}
\end{figure}

\begin{figure}
\centering
   \includegraphics[width=\columnwidth]{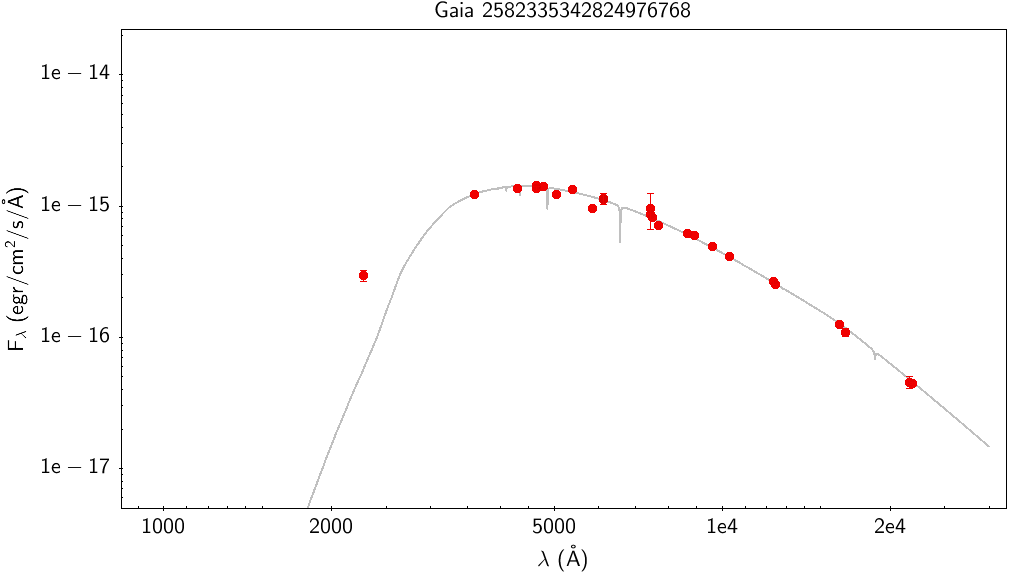}
   \includegraphics[width=\columnwidth]{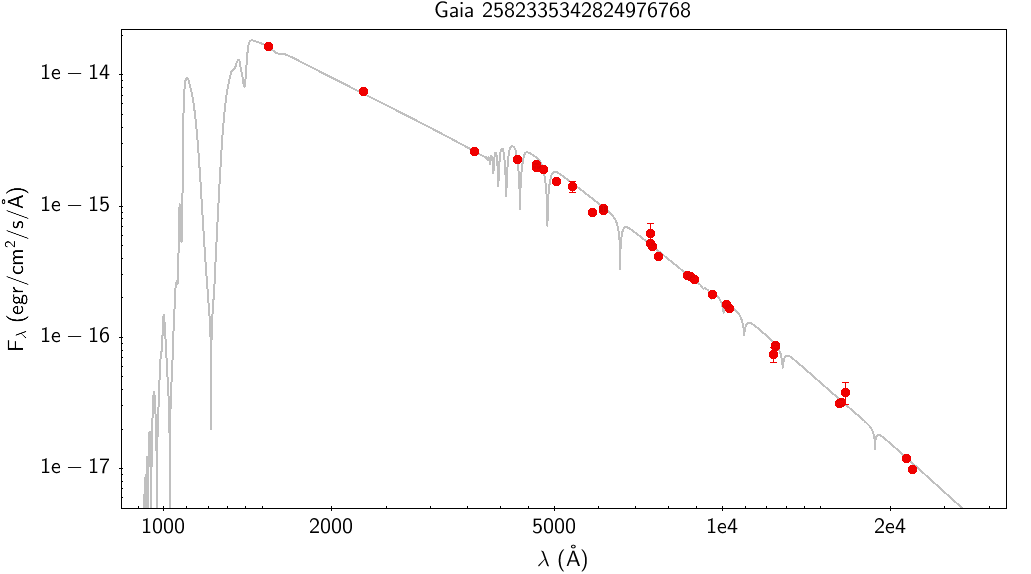}
\caption{Examples of observational SEDs and the synthetic spectra that best fit the observations. The observational photometric points and their errors are shown in red and the best-fitted models in gray. The upper panel corresponds to a synthetic spectrum of T$_{eff}$\,=\,6,250 K and log g\,=\,8.75 dex, and the lower to one of T$_{eff}$\,=\,14,000 K and log g\,=\,8.25 dex.} 
\label{f-SEDS}
\end{figure}

\subsection{The SDSS control sample}

In order to evaluate both the accuracy of the fitting method and the reliability of the derived effective temperature values, we selected $\sim$4,000 CO-core DA white dwarfs from the SDSS white dwarf catalogue of \citet{Anguiano17} for which accurate spectroscopic effective temperature values are available (i.e. a relative error below 10\%), and fitted them using VOSA. The SDSS sample is located at a much larger distance than our {\it Gaia} sample, with the peak of the distribution at $\sim$200\,pc and a decreasing long tail towards 1500 pc \citep{Anguiano17}. Thus, the effect of the interstellar extinction in the estimation of the effective temperature will be much larger for the SDSS sample than for our {\it Gaia} sample. Consequently, we used the values of the interstellar extinction by \citet{Schlafly+Finkbeiner11} for the SDSS sample.

The VOSA results obtained are compared to the spectroscopic values in Fig.\,\ref{f-sdssvosa}. We plotted those objects with good SED fitting and with derived parameter values not falling at the edge of the model atmosphere grid (i.e. Vgfb\footnote{Vgfb: Modified reduced $\chi^{2}$, calculated by forcing $\sigma(F_{obs})$ to be larger than $0.1\times F_{obs}$, where $\sigma(F_{obs})$ is the error in the observed flux ($F_{obs}$). This can be useful if the photometric errors of any of the catalogues used to build the SED are underestimated. Vgfb smaller than 10--15 is often perceived as a good fit.}\,$<$\,15 and 6.5\,$<$\,log g\,$<$\,9.5).

Figure\,\ref{f-sdssvosa} clearly reveals a correlation between the spectroscopic SDSS and the photometric VOSA effective temperatures for the vast majority of WDs. This correlation mostly followed the 1:1 relation (dashed line), with a light deviation at the extremes of the temperature range. 

A closer inspection of Fig.\,\ref{f-sdssvosa} reveals an additional feature at spectroscopic T$_\mathrm{eff} \sim 10,000-15,000$\,K. In particular, for some white dwarfs the VOSA values of effective temperature are considerable higher than the spectroscopic ones. The 10,000--15,000\,K temperature range is known for the maximum occurrence of the H$\beta$ equivalent width (the exact value depends on the surface gravity of the white dwarf; \citealt{Rebassa07}). Therefore, the effective temperature (and surface gravity) values determined from Balmer line profile fits are subject to a degeneracy, i.e. fits of similar quality can be achieved on either side of the temperature at which the maximum equivalent width is occurring. These are often referred to as the 'hot' and 'cold' solutions. Hence, the most likely reason for the spectroscopic fits providing lower values of effective temperature is that the spectroscopic solution is the wrong one.

Based on Fig.\,\ref{f-sdssvosa}, we concluded VOSA provides reliable values of effective temperature for the great majority of white dwarfs.

\subsection{VOSA Results}
\label{VOSAresults}

We used VOSA to build the observational SED and to fit it to the DA white dwarf model spectra of \cite{Koester10} to each {\it Gaia} source of our catalogue. Of the 72,178 CO-core and the 1,043 ONe-core, we obtained the effective temperature and the luminosity for 65,955 ($\sim$91\%) and 938 ($\sim$90\%) sources, respectively, most of them (42,415 and 509) with a reliable fit (i.e. Vgfb\,$<$\,15 and log g different than 6.5 or 9.5, the extremes of the model grid). The rest of sources had not enough photometric points for the fitting.

In order to reduce the possible contamination due to photometric errors in the BP and RP bands, but retaining a representative number of cool white dwarfs in our sample, we imposed the colour cut $(G_{\rm BP}-G_{\rm RP})<0.80$, corresponding to an effective theoretical temperature of $\sim$\,6,000 K. However, it was expected that some cooler white dwarfs would have passed our search colour criteria because of the observational errors. Thus, we used the whole range of temperatures provided by the model atmosphere grid (5,000 K -- 80,000 K), finding 243 CO-core white dwarfs with estimated effective temperatures between 5,000 and 6,000 K.

For illustrative purposes, in Fig. \ref{f-SEDS} we show the observational SEDs built from VO photometric catalogues together with the synthetic spectrum that best fits the data for two {\it Gaia} sources of our catalogue of white dwarfs, one cool and one hot. Both synthetic spectra reproduce very well the observational SEDs. Only the UV region for the cool white dwarf is worst reproduced by the model, however this represents a minor part of the total energy output. Thus, we conclude that VOSA provides reliable values of effective temperatures and bolometric fluxes. 

Among the 8,343 CO-core and 212 ONe-core white dwarfs located within 100\,pc from the Sun, the VOSA SED fitting provided reliable results for 52\%, i.e. 4,317 CO-core and 88 ONe-core white dwarfs. Taking into account that we expect $\sim$44\% of all white dwarfs within 100\,pc to have temperatures between 6,000 K and 80,000 K (see Table\,\ref{t:table1}), this implies that we obtained reliable effective temperature and luminosity values for $\sim$23\% of all expected white dwarfs existing within 100\,pc. It is worth noting here that these percentages refer to the completeness of the sample and not to the accuracy of the fittings. Although the final fraction of objects with reliable parameters may be considered small, it has been extracted from a nearly complete sample. Besides, the accurate procedure of obtaining reliable parameters from VOSA SED fitting brings us to consider that our final sample of white dwarfs with well determined parameters is representative of the complete sample within the effective temperature range under study.


\section{The physical properties of the 100\,pc {\it Gaia} white dwarfs}
\label{phys-prop}

\begin{figure*}
   \includegraphics[trim=5mm 30mm 5mm 20mm, clip,width=1.8\columnwidth]{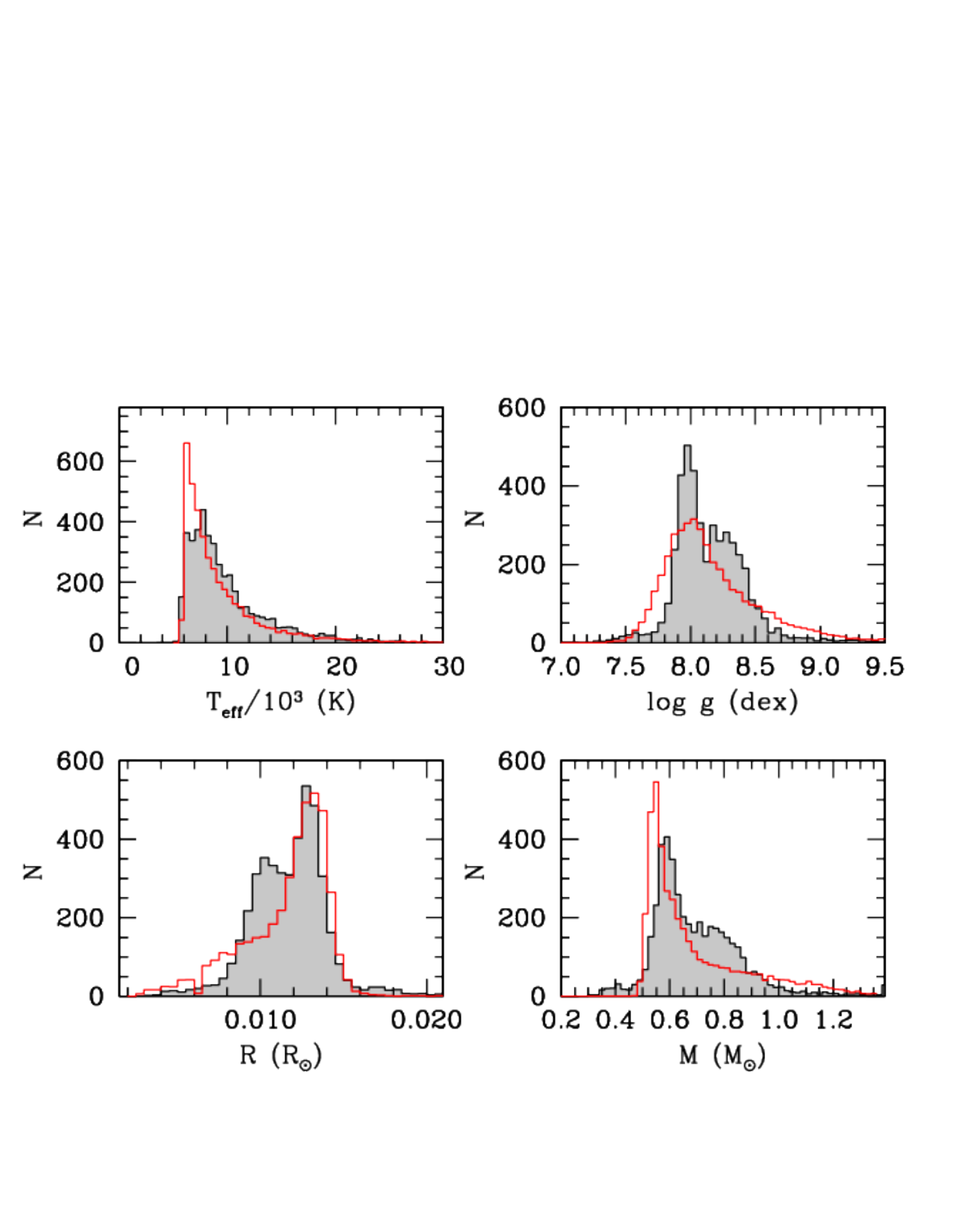}
\caption{Effective temperature (top-left), radius (bottom-left), surface gravity (top-right) and mass (bottom-right) distributions for {\it Gaia} white dwarfs (grey histograms) within 100\,pc and with effective temperatures between 5,000 K and 80,000 K. Only sources with good SED fitting were considered (see Section \ref{VOSAresults}). For comparative purposes we also show the corresponding distributions arising from our synthetic population models (red solid lines). The observed {\it Gaia}  100\,pc sample  contains a significant number of massive ($M \sim 0.8$ M$_{\odot}$) white dwarfs and a lack of cool objects ($T_{\rm eff} \la$ 8,000 K). See text for details.}
\label{f-param}
\end{figure*}

The analysis of the photometric properties of the {\it Gaia} white dwarf catalogue exposed so far permitted us to study some of the most relevant properties of this population. As previously stated, our estimates for the 100\,pc sample indicate that we can recover nearly all white dwarfs within effective temperatures between 6,000 K and 80,000 K. At the same time, this sub-sample represents a $\sim$44\% of all the white dwarf population within 100\,pc.

First of all, we analysed the effective temperature distribution of our sample of white dwarfs within 100\,pc, as obtained from the VOSA fits. This is illustrated in the top left panel of Fig.\,\ref{f-param} (grey histogram). For comparative purposes we also show the corresponding distribution arising from our synthetic population models (red solid lines) normalized to the same number of objects as the observed sample. Inspection of the referred panel reveals a clear concentration of white dwarfs at the coolest bins (T$_\mathrm{eff} \sim$ 8,000 K) followed by an exponential decline towards hotter temperatures, as typically expected for the white dwarf cooling process. In general terms, the observed distribution perfectly agrees with the simulated one. However, there is clear a lack of observed white dwarfs at the coolest temperature range (T$_\mathrm{eff} \la$ 8,000 K). As previously explained in Section\,\ref{query} and shown in Fig. \ref{f-hr-excess}, the cut in the {\it phot\_bp\_rp\_excess\_factor} eliminated those objects with large  photometric errors in the BP and RP bands, specially for redder colours, i.e. cooler effective temperatures. Additionally, objects with bad VOSA stellar parameter estimates, which were eliminated from the present study, were in a greater proportion also redder white dwarfs (see Section\,\ref{results}). As a consequence, some of these objects were lost from our 100\,pc sample, in particular those with the lowest temperatures.

Secondly, we used the effective temperatures and the luminosities provided by VOSA to derive radii for our white dwarfs from the well known relation $L=4 \pi R^{2} \sigma T^{4}$, where $L$ is the luminosity, $T$ is the effective temperature, $R$ is the radius and $\sigma$ is the Stefan-Boltzmann constant. The corresponding radius histogram is shown in the bottom-left panel of Fig.\,\ref{f-param}. As can be seen, the most remarkable feature is a bimodal-like distribution with peaks at $\sim$0.010 and 0.013 $R_{\odot}$, being the lowest peak not predicted by our synthetic population models. It is worth mentioning that this low-radius peak presented in the observed distribution is a clear indicative of the existence of a significant population of massive white dwarfs. 

Finally, we interpolated the effective temperature and radius values of our {\it Gaia} objects in the evolutionary sequences for DA white dwarfs of \citet{Renedo2010} to obtain their surface gravities and masses. The corresponding distributions are displayed in the top-right and bottom-right panels (grey histograms) of Fig.\,\ref{f-param}, respectively. As expected, the bimodal-like radius distribution translates into a bimodal pattern for both the surface gravity and the mass distributions, indeed revealing the existence of a significant population of massive (M$\sim0.8 M_{\odot}$) white dwarfs. As previously indicated in Section \ref{results}, the magnitude-colour diagram of the {\it Gaia} 100\,pc sample presents a characteristic bifurcation (see Fig. \ref{f-hrcat}) which now is put into manifest by the presence of an excess of massive white dwarfs. The existence of such a population, although not predicted by our models, was suggested by previous studies analysing both magnitude-limited samples to which one needs to apply a volume correction \citep{Liebert05, Rebassa15-1, Rebassa15-2}, as well as by smaller-size volume-limited white dwarf compilations \citep{Giammichele12, Limoges15}. 

Possible discrepancies between DA and DB cooling tracks do not seem to be plausible explanations for this observed bifurcation and excess of massive white dwarfs. As shown in section \ref{GaiaHR}, our estimates indicate that around $\approx$\,30\,--\,40\% of the objects belonging to the bifurcation region may be DB white dwarfs. On the other hand, since we did not have a priori information about the atmospheric chemical characteristics of the white dwarfs found, our stellar parameter estimates were based on the assumption that all white dwarfs were DAs, which was true for most of them. Consequently, we used DA cooling sequences for the whole sample. Thus, in principle, we expected unreliable stellar parameter determinations (specially surface gravity and mass values) for up to 1/5 of our objects. However, as we found out cross-correlating our catalogue with the SDSS DB sample of \citet{Koester2015}, VOSA provided bad fits (Vgfb\,>\,15) for $\sim$70\% of DB white dwarfs. This implies that, by excluding objects with bad VOSA fits, the  contribution of DB white dwarfs to our final parameter-determined sample should be smaller than $\approx$\,6\%. This small percentage hardly seems to explain the over-abundance of massive white dwarfs.

Several hypothesis for the origin of this observed high-mass excess have raised so far. The most extended explanation grants its origin to white dwarf binary mergers \citep{Kilic2018}. Recently,  \cite{Maoz18} claimed that up to a $\sim$10 per cent of the observed single white dwarfs are expected to be the result of binary white dwarf mergers, thus significantly contributing to the massive content of these objects. Alternatively, binary scenarios such as the merger of the degenerate core of a giant star with a main-sequence or a white dwarf companion \citep{Rebassa15-2} have also been proposed. Other possibilities, which are not a priori disposable, also exist. For instance, a recent burst of star formation, as argued by  \cite{Torres2016} to explain the local 40 pc white dwarf luminosity function, may also lead to an excess of massive white dwarfs.

Moreover, the initial-to-final mass relationship for white dwarfs was proved to play a key role in the final white dwarf mass distribution \citep[e.g.][]{Williams2004,Catala2008,Meng2008,Kalirai2008,Andrews2015,ElBadry2018}. Even the origin of a significant fraction of massive white dwarfs has been related to the existence of high-field magnetic white dwarfs \citep{GBerro2012}. Anyhow, it is beyond the scope of the present paper to provide a definitive explanation to this issue. Here, we just confirm the existence of a real excess of massive white dwarfs in the 100\,pc sample that can not be attributed to any observational or statistical bias.

Our observational mass distribution also reveals the typical and more dominant peak at $\sim$0.6 $M_{\odot}$ (equivalent to $\log$\,g $\sim$8 dex). A closer look at Fig.\,\ref{f-param} reveals that this main observed peak is slightly shifted towards a higher value than that predicted by our population models, which is closer to $\sim$0.55 $M_{\odot}$. We can point out some factors that may contribute to the observed shift in the peak of the mass distribution. Among these, we can cite (i) a systematic error in the mass estimate; (ii) a steeper slope of the IFMR would promote more massive white dwarfs, shifting in consequence the peak of the distribution; (iii) the contribution of white dwarfs coming from mergers. Any of these factors or a combination of them is expected to produce a displacement in the peak distribution. We postpone a thorough analysis of the mass distribution of {\it Gaia} white dwarfs to a forthcoming paper.

Finally, our results also shown the existence of a small fraction of He-core white dwarfs ($\sim$0.45 $M_{\odot}$) and/or unresolved double-degenerate objects (see the discussion in Section \ref{results}).


\section{Conclusions}
\label{conclusion}

The Data Release 2 of {\it Gaia} has provided the scientific community with an unprecedented collection, both in quality and quantity, of accurate photometric and astrometric data. The work presented here represents a major effort to extract a comprehensive up-to-date catalogue of white dwarfs from {\it Gaia} DR2 data and to derive their more relevant parameters.

With the aid of a detailed Monte Carlo simulator we built a thorough population synthesis model of the thin disc, thick disc, and halo white dwarf population of our Galaxy. Our synthetic models incorporate updated evolutionary sequences for both CO-core and ONe-core white dwarfs, as well as hydrogen-rich and hydrogen-deficient atmospheres. Additionally, we included in our models a complete treatment of the different selection criteria and photometric and astrometric errors predicted by {\it Gaia} performances. This permitted us to define the regions within the {\it Gaia} HR-diagram where we expected to identify white dwarfs, as well as to predict its completeness and to estimate its possible contamination.

Our search provided the largest white dwarf catalogue to date, which includes more than 73,000 objects with accurate {\it Gaia} parallax ($\pi/\sigma_{\pi}$\,$>$\,10) and photometry ($F_{\rm G}/\sigma_{F_{\rm G}}$\,$>$\,50, $F_{\rm BP}/\sigma_{F_{\rm BP}}$\,$>$\,10, $F_{\rm RP}/\sigma_{F_{\rm RP}}$\,$>$\,10). From this sample, we identified 8,343 CO-core and 212 ONe-core white dwarf candidates within 100\,pc in the range of effective temperatures from 5,000 K to 80,000 K, from which we derive a white dwarf space density of $4.9\pm0.4\times10^{-3}\,{\rm pc^{-3}}$, consistent with previous studies. In turn, our population synthesis models indicated that the completeness of the white dwarf population accessible by {\it Gaia} and with accurate parallax measurements is nearly complete, $94\%$, up to 100\,pc from the Sun. Besides, our analysis also showed that this distance represents the maximum distance that {\it Gaia} can achieve for a volume-limited sample of white dwarfs. Hence, we remark that our sub-sample up to 100\,pc and within the range of temperatures before mentioned represents the most updated highly-complete volume-limited sample of white dwarfs.

The {\it Gaia} HR-diagram obtained for our 100\,pc sample revealed a bifurcation trend at around  0\,$\lappr$\,$G_{\rm BP}-G_{\rm RP}$\,$\lappr$\,0.3. This bifurcation is not predicted by actual, neither cooling nor galactic, white dwarf models. Possible discrepancies between DA or DB white dwarf cooling tracks are not either able to explain the observed bifurcation for the range of colours considered. Our analysis also reveals that the lower branch of the observed bifurcation is formed by no more than a $30-40\%$ of DB white dwarfs. Furthermore, the observed bifurcation is not attributable to any observational bias or statistical artifact and it is consequently a real characteristic of the 100\,pc white dwarf population. Additionally, our analysis reveals the contamination from other astronomical sources is expected to be very low ($\lappr$\,1\%) and mainly due to cataclysmic variables and subdwarf B stars. Finally, the limitation in the excess of flux colour -- necessary to be applied to our sample in order to avoid highly crowed contaminated areas -- is found to be  only marginally important for low temperature objects, $T_\mathrm{eff} \lappr\,$8,000 K.

Estimating stellar parameters is required to perform any accurate  study. In this sense, we used an updated grid of hydrogen-rich white dwarf model atmosphere spectra in the Virtual Observatory tool VOSA to fit the observational spectral energy distributions of our sources, which were built from photometric catalogues from the UV to the NIR publicly available at the VO. We demonstrated the fits provide reliable white dwarf effective temperature and luminosity values, which were derived for $\sim$59\% of all {\it Gaia} white dwarfs in the considered colour range ($-\,0.52<(G_{\rm BP}-G_{\rm RP})<0.80$). We subsequently used the VOSA luminosities and effective temperatures to derive the radii of our white dwarfs. Interpolating the effective temperature and radius values on the cooling sequences of hydrogen-rich white dwarfs, we finally derived their surface gravities and masses.

The effective temperature distribution displayed a prominent population of cool white dwarfs peaking at $T_\mathrm{eff}\sim$8,000\,K and a paucity of hot objects ($T_\mathrm{eff}>$20,000\,K). This clearly demonstrates the majority of white dwarfs in the solar neighbourhood are cool, an expected result that previous magnitude-limited samples failed to reproduce. Moreover, we identified bimodal-like distributions of radius, surface gravity and mass, the three of them containing the ``canonical'' ($M \sim 0.6\,M_{\odot}$) population of white dwarfs, as well as a significant population of more massive ($M \sim 0.8 M_{\odot}$) white dwarfs. This excess of massive white dwarfs is clearly related to the observed bifurcation of the HR-diagram and, as we have shown in this paper, it can not be entirely attributed to the existence of helium-rich atmosphere white dwarfs. Its origin remains uncertain and ensures amazing forthcoming studies. In this sense, deeper analyses including identified white dwarf spectral types or the Galactic component classification, among other issues, are required to extract the maximum information from the Gaia white dwarf catalogue presented here.

\section*{Acknowledgements}

We are greatly indebted to Detlev Koester for providing us with his white dwarf model atmosphere spectra and for helpful discussions. This work  was partially supported by the MINECO grant AYA\-2017-86274-P and the Ram\'on y Cajal programme RYC-2016-20254 and by the AGAUR (SGR-661/2017). F.J.E. acknowledges financial support from ASTERICS project (ID:653477, H2020-EU.1.4.1.1. - Developing new world-class research infrastructures).

This work has made use of data from the European Space Agency (ESA) mission {\it Gaia} (\url{https://www.cosmos.esa.int/gaia}), processed by the {\it Gaia} Data Processing and Analysis Consortium (DPAC, \url{https://www.cosmos.esa.int/web/gaia/dpac/consortium}). Funding for the DPAC has been provided by national institutions, in particular the institutions participating in the {\it {\it Gaia}} Multilateral Agreement. This publication makes use of VOSA, developed under the Spanish Virtual Observatory project supported from the Spanish MINECO through grant AyA2017-84089. We extensively made used of Topcat \citep{Taylor05}. his research has made use of the VizieR catalogue access tool, CDS, Strasbourg, France. We acknowledge use of the ADS bibliographic services.

\bibliographystyle{mnras}
\bibliography{references} 

\begin{thebibliography}{}
\makeatletter
\relax
\def\mn@urlcharsother{\let\do\@makeother \do\$\do\&\do\#\do\^\do\_\do\%\do\~}
\def\mn@doi{\begingroup\mn@urlcharsother \@ifnextchar [ {\mn@doi@}
  {\mn@doi@[]}}
\def\mn@doi@[#1]#2{\def\@tempa{#1}\ifx\@tempa\@empty \href
  {http://dx.doi.org/#2} {doi:#2}\else \href {http://dx.doi.org/#2} {#1}\fi
  \endgroup}
\def\mn@eprint#1#2{\mn@eprint@#1:#2::\@nil}
\def\mn@eprint@arXiv#1{\href {http://arxiv.org/abs/#1} {{\tt arXiv:#1}}}
\def\mn@eprint@dblp#1{\href {http://dblp.uni-trier.de/rec/bibtex/#1.xml}
  {dblp:#1}}
\def\mn@eprint@#1:#2:#3:#4\@nil{\def\@tempa {#1}\def\@tempb {#2}\def\@tempc
  {#3}\ifx \@tempc \@empty \let \@tempc \@tempb \let \@tempb \@tempa \fi \ifx
  \@tempb \@empty \def\@tempb {arXiv}\fi \@ifundefined
  {mn@eprint@\@tempb}{\@tempb:\@tempc}{\expandafter \expandafter \csname
  mn@eprint@\@tempb\endcsname \expandafter{\@tempc}}}

\bibitem[\protect\citeauthoryear{{Alam} et~al.,}{{Alam} et~al.}{2015}]{Alam15}
{Alam} S.,  et~al., 2015, \mn@doi [\apjs] {10.1088/0067-0049/219/1/12}, \href
  {http://cdsads.u-strasbg.fr/abs/2015ApJS..219...12A} {219, 12}

\bibitem[\protect\citeauthoryear{{Althaus}, {Serenelli}, {Panei},
  {C{\'o}rsico}, {Garc{\'{\i}}a-Berro}  \& {Sc{\'o}ccola}}{{Althaus}
  et~al.}{2005}]{Althaus2005}
{Althaus} L.~G.,  {Serenelli} A.~M.,  {Panei} J.~A.,  {C{\'o}rsico} A.~H.,
  {Garc{\'{\i}}a-Berro} E.,   {Sc{\'o}ccola} C.~G.,  2005, \mn@doi [\aap]
  {10.1051/0004-6361:20041965}, \href
  {http://adsabs.harvard.edu/abs/2005A%26A...435..631A} {435, 631}

\bibitem[\protect\citeauthoryear{{Althaus}, {Garc{\'{\i}}a-Berro}, {Isern},
  {C{\'o}rsico}  \& {Rohrmann}}{{Althaus} et~al.}{2007}]{Althaus2007}
{Althaus} L.~G.,  {Garc{\'{\i}}a-Berro} E.,  {Isern} J.,  {C{\'o}rsico} A.~H.,
   {Rohrmann} R.~D.,  2007, \mn@doi [\aap] {10.1051/0004-6361:20066059}, \href
  {http://cdsads.u-strasbg.fr/abs/2007A%26A...465..249A} {465, 249}

\bibitem[\protect\citeauthoryear{{Althaus}, {C{\'o}rsico}, {Isern}  \&
  {Garc{\'{\i}}a-Berro}}{{Althaus} et~al.}{2010}]{Althaus2010a}
{Althaus} L.~G.,  {C{\'o}rsico} A.~H.,  {Isern} J.,   {Garc{\'{\i}}a-Berro} E.,
   2010, \mn@doi [\aapr] {10.1007/s00159-010-0033-1}, \href
  {http://cdsads.u-strasbg.fr/abs/2010A%26ARv..18..471A} {18, 471}

\bibitem[\protect\citeauthoryear{{Althaus}, {Camisassa}, {Miller Bertolami},
  {C{\'o}rsico}  \& {Garc{\'{\i}}a-Berro}}{{Althaus}
  et~al.}{2015}]{Althaus2015}
{Althaus} L.~G.,  {Camisassa} M.~E.,  {Miller Bertolami} M.~M.,  {C{\'o}rsico}
  A.~H.,   {Garc{\'{\i}}a-Berro} E.,  2015, \mn@doi [\aap]
  {10.1051/0004-6361/201424922}, \href
  {http://adsabs.harvard.edu/abs/2015A%26A...576A...9A} {576, A9}

\bibitem[\protect\citeauthoryear{{Andrews}, {Ag{\"u}eros}, {Gianninas},
  {Kilic}, {Dhital}  \& {Anderson}}{{Andrews} et~al.}{2015}]{Andrews2015}
{Andrews} J.~J.,  {Ag{\"u}eros} M.~A.,  {Gianninas} A.,  {Kilic} M.,  {Dhital}
  S.,   {Anderson} S.~F.,  2015, \mn@doi [\apj] {10.1088/0004-637X/815/1/63},
  \href {http://adsabs.harvard.edu/abs/2015ApJ...815...63A} {815, 63}

\bibitem[\protect\citeauthoryear{{Anguiano}, {Rebassa-Mansergas},
  {Garc{\'{\i}}a-Berro}, {Torres}, {Freeman}  \& {Zwitter}}{{Anguiano}
  et~al.}{2017}]{Anguiano17}
{Anguiano} B.,  {Rebassa-Mansergas} A.,  {Garc{\'{\i}}a-Berro} E.,  {Torres}
  S.,  {Freeman} K.~C.,   {Zwitter} T.,  2017, \mn@doi [\mnras]
  {10.1093/mnras/stx796}, \href
  {http://adsabs.harvard.edu/abs/2017MNRAS.469.2102A} {469, 2102}

\bibitem[\protect\citeauthoryear{{Barstow} et~al.,}{{Barstow}
  et~al.}{2014}]{Barstow2014}
{Barstow} M.~A.,  et~al., 2014, preprint, \href
  {http://adsabs.harvard.edu/abs/2014arXiv1407.6163B} {} (\mn@eprint {arXiv}
  {1407.6163})

\bibitem[\protect\citeauthoryear{{Bayo}, {Rodrigo}, {Barrado y Navascu{\'e}s},
  {Solano}, {Guti{\'e}rrez}, {Morales-Calder{\'o}n}  \& {Allard}}{{Bayo}
  et~al.}{2008}]{Bayo08}
{Bayo} A.,  {Rodrigo} C.,  {Barrado y Navascu{\'e}s} D.,  {Solano} E.,
  {Guti{\'e}rrez} R.,  {Morales-Calder{\'o}n} M.,   {Allard} F.,  2008, \mn@doi
  [\aap] {10.1051/0004-6361:200810395}, \href
  {http://adsabs.harvard.edu/abs/2008A%26A...492..277B} {492, 277}

\bibitem[\protect\citeauthoryear{{Bergeron} et~al.,}{{Bergeron}
  et~al.}{2011}]{Bergeron2011}
{Bergeron} P.,  et~al., 2011, \mn@doi [\apj] {10.1088/0004-637X/737/1/28},
  \href {http://adsabs.harvard.edu/abs/2011ApJ...737...28B} {737, 28}

\bibitem[\protect\citeauthoryear{{Bianchi} \& {GALEX Team}}{{Bianchi} \& {GALEX
  Team}}{2000}]{Bianchi00}
{Bianchi} L.,  {GALEX Team} 2000, \memsai, \href
  {http://adsabs.harvard.edu/abs/2000MmSAI..71.1123B} {71, 1123}

\bibitem[\protect\citeauthoryear{{Bland-Hawthorn} \&
  {Gerhard}}{{Bland-Hawthorn} \& {Gerhard}}{2016}]{Bland-Hawthorn2016}
{Bland-Hawthorn} J.,  {Gerhard} O.,  2016, \mn@doi [\araa]
  {10.1146/annurev-astro-081915-023441}, \href
  {http://adsabs.harvard.edu/abs/2016ARA%26A..54..529B} {54, 529}

\bibitem[\protect\citeauthoryear{{Bradley}}{{Bradley}}{1998}]{Bradley1998}
{Bradley} P.~A.,  1998, \mn@doi [\apjs] {10.1086/313102}, \href
  {http://adsabs.harvard.edu/abs/1998ApJS..116..307B} {116, 307}

\bibitem[\protect\citeauthoryear{{Calamida} et~al.,}{{Calamida}
  et~al.}{2008}]{Calamida2008}
{Calamida} A.,  et~al., 2008, \mn@doi [\apjl] {10.1086/527436}, \href
  {http://adsabs.harvard.edu/abs/2008ApJ...673L..29C} {673, L29}

\bibitem[\protect\citeauthoryear{{Calamida} et~al.,}{{Calamida}
  et~al.}{2014}]{Calamida2014}
{Calamida} A.,  et~al., 2014, \mn@doi [\apj] {10.1088/0004-637X/790/2/164},
  \href {http://adsabs.harvard.edu/abs/2014ApJ...790..164C} {790, 164}

\bibitem[\protect\citeauthoryear{{Camisassa}, {Althaus}, {Rohrmann},
  {Garc{\'{\i}}a-Berro}, {Torres}, {C{\'o}rsico}  \& {Wachlin}}{{Camisassa}
  et~al.}{2017}]{Camisasa2017}
{Camisassa} M.~E.,  {Althaus} L.~G.,  {Rohrmann} R.~D.,  {Garc{\'{\i}}a-Berro}
  E.,  {Torres} S.,  {C{\'o}rsico} A.~H.,   {Wachlin} F.~C.,  2017, \mn@doi
  [\apj] {10.3847/1538-4357/aa6797}, \href
  {http://adsabs.harvard.edu/abs/2017ApJ...839...11C} {839, 11}

\bibitem[\protect\citeauthoryear{{Castanheira} \& {Kepler}}{{Castanheira} \&
  {Kepler}}{2008}]{Castanheira2008}
{Castanheira} B.~G.,  {Kepler} S.~O.,  2008, \mn@doi [\mnras]
  {10.1111/j.1365-2966.2008.12851.x}, \href
  {http://adsabs.harvard.edu/abs/2008MNRAS.385..430C} {385, 430}

\bibitem[\protect\citeauthoryear{{Castellani}, {Cignoni}, {Degl'Innocenti},
  {Petroni}  \& {Prada Moroni}}{{Castellani} et~al.}{2002}]{Castellani2002}
{Castellani} V.,  {Cignoni} M.,  {Degl'Innocenti} S.,  {Petroni} S.,   {Prada
  Moroni} P.~G.,  2002, \mn@doi [\mnras] {10.1046/j.1365-8711.2002.05461.x},
  \href {http://adsabs.harvard.edu/abs/2002MNRAS.334...69C} {334, 69}

\bibitem[\protect\citeauthoryear{{Catal{\'a}n}, {Isern}, {Garc{\'{\i}}a-Berro}
  \& {Ribas}}{{Catal{\'a}n} et~al.}{2008}]{Catala2008}
{Catal{\'a}n} S.,  {Isern} J.,  {Garc{\'{\i}}a-Berro} E.,   {Ribas} I.,  2008,
  \mn@doi [\mnras] {10.1111/j.1365-2966.2008.13356.x}, \href
  {http://adsabs.harvard.edu/abs/2008MNRAS.387.1693C} {387, 1693}

\bibitem[\protect\citeauthoryear{{Chambers} et~al.,}{{Chambers}
  et~al.}{2016}]{Chambers16}
{Chambers} K.~C.,  et~al., 2016, preprint, \href
  {http://cdsads.u-strasbg.fr/abs/2016arXiv161205560C} {} (\mn@eprint {arXiv}
  {1612.05560})

\bibitem[\protect\citeauthoryear{{Cojocaru}, {Torres}, {Althaus}, {Isern}  \&
  {Garc{\'{\i}}a-Berro}}{{Cojocaru} et~al.}{2015}]{Cojocaru2015}
{Cojocaru} R.,  {Torres} S.,  {Althaus} L.~G.,  {Isern} J.,
  {Garc{\'{\i}}a-Berro} E.,  2015, \mn@doi [\aap]
  {10.1051/0004-6361/201526550}, \href
  {http://adsabs.harvard.edu/abs/2015A%26A...581A.108C} {581, A108}

\bibitem[\protect\citeauthoryear{{Cross} et~al.,}{{Cross}
  et~al.}{2012}]{Cross12}
{Cross} N.~J.~G.,  et~al., 2012, \mn@doi [\aap] {10.1051/0004-6361/201219505},
  \href {http://adsabs.harvard.edu/abs/2012A%26A...548A.119C} {548, A119}

\bibitem[\protect\citeauthoryear{{Dark Energy Survey Collaboration}
  et~al.,}{{Dark Energy Survey Collaboration} et~al.}{2016}]{DESC16}
{Dark Energy Survey Collaboration} et~al., 2016, \mn@doi [\mnras]
  {10.1093/mnras/stw641}, \href
  {http://adsabs.harvard.edu/abs/2016MNRAS.460.1270D} {460, 1270}

\bibitem[\protect\citeauthoryear{{Dreiner}, {Fortin}, {Isern}  \&
  {Ubaldi}}{{Dreiner} et~al.}{2013}]{Dreiner2013}
{Dreiner} H.~K.,  {Fortin} J.-F.,  {Isern} J.,   {Ubaldi} L.,  2013, \mn@doi
  [\prd] {10.1103/PhysRevD.88.043517}, \href
  {http://adsabs.harvard.edu/abs/2013PhRvD..88d3517D} {88, 043517}

\bibitem[\protect\citeauthoryear{{Eisenstein} et~al.,}{{Eisenstein}
  et~al.}{2006}]{Eisenstein2006}
{Eisenstein} D.~J.,  et~al., 2006, \mn@doi [\apjs] {10.1086/507110}, \href
  {http://adsabs.harvard.edu/abs/2006ApJS..167...40E} {167, 40}

\bibitem[\protect\citeauthoryear{{El-Badry}, {Rix}  \& {Weisz}}{{El-Badry}
  et~al.}{2018}]{ElBadry2018}
{El-Badry} K.,  {Rix} H.-W.,   {Weisz} D.~R.,  2018, preprint, \href
  {http://adsabs.harvard.edu/abs/2018arXiv180505849E} {} (\mn@eprint {arXiv}
  {1805.05849})

\bibitem[\protect\citeauthoryear{{Evans}, {Irwin}  \& {Helmer}}{{Evans}
  et~al.}{2002}]{Evans02}
{Evans} D.~W.,  {Irwin} M.~J.,   {Helmer} L.,  2002, \mn@doi [\aap]
  {10.1051/0004-6361:20021285}, \href
  {http://adsabs.harvard.edu/abs/2002A%26A...395..347E} {395, 347}

\bibitem[\protect\citeauthoryear{{Evans} et~al.,}{{Evans}
  et~al.}{2018}]{Evans18}
{Evans} D.~W.,  et~al., 2018, preprint, \href
  {http://adsabs.harvard.edu/abs/2018arXiv180409368E} {} (\mn@eprint {arXiv}
  {1804.09368})

\bibitem[\protect\citeauthoryear{{Gaia Collaboration} et~al.,}{{Gaia
  Collaboration} et~al.}{2018a}]{Babusiaux2018}
{Gaia Collaboration} et~al., 2018a, preprint, \href
  {http://adsabs.harvard.edu/abs/2018arXiv180409378G} {} (\mn@eprint {arXiv}
  {1804.09378})

\bibitem[\protect\citeauthoryear{{Gaia Collaboration}, {Brown}, {Vallenari},
  {Prusti}, {de Bruijne}, {Babusiaux}  \& {Bailer-Jones}}{{Gaia Collaboration}
  et~al.}{2018b}]{Brown18}
{Gaia Collaboration} {Brown} A.~G.~A.,  {Vallenari} A.,  {Prusti} T.,  {de
  Bruijne} J.~H.~J.,  {Babusiaux} C.,   {Bailer-Jones} C.~A.~L.,  2018b,
  preprint, \href {http://cdsads.u-strasbg.fr/abs/2018arXiv180409365G} {}
  (\mn@eprint {arXiv} {1804.09365})

\bibitem[\protect\citeauthoryear{{Garcia-Berro} \& {Iben}}{{Garcia-Berro} \&
  {Iben}}{1994}]{GBerro1994}
{Garcia-Berro} E.,  {Iben} I.,  1994, \mn@doi [\apj] {10.1086/174729}, \href
  {http://adsabs.harvard.edu/abs/1994ApJ...434..306G} {434, 306}

\bibitem[\protect\citeauthoryear{{Garc{\'{\i}}a-Berro} \&
  {Oswalt}}{{Garc{\'{\i}}a-Berro} \& {Oswalt}}{2016}]{GBerroOswalt2016}
{Garc{\'{\i}}a-Berro} E.,  {Oswalt} T.~D.,  2016, \mn@doi [\nar]
  {10.1016/j.newar.2016.08.001}, \href
  {http://adsabs.harvard.edu/abs/2016NewAR..72....1G} {72, 1}

\bibitem[\protect\citeauthoryear{{Garcia-Berro}, {Hernanz}, {Isern}  \&
  {Mochkovitch}}{{Garcia-Berro} et~al.}{1988}]{GB88b}
{Garcia-Berro} E.,  {Hernanz} M.,  {Isern} J.,   {Mochkovitch} R.,  1988,
  \mn@doi [\nat] {10.1038/333642a0}, \href
  {http://adsabs.harvard.edu/abs/1988Natur.333..642G} {333, 642}

\bibitem[\protect\citeauthoryear{{Garcia-Berro}, {Hernanz}, {Isern}  \&
  {Mochkovitch}}{{Garcia-Berro} et~al.}{1995}]{GBerro1995}
{Garcia-Berro} E.,  {Hernanz} M.,  {Isern} J.,   {Mochkovitch} R.,  1995,
  \mn@doi [\mnras] {10.1093/mnras/277.3.801}, \href
  {http://cdsads.u-strasbg.fr/abs/1995MNRAS.277..801G} {277, 801}

\bibitem[\protect\citeauthoryear{{Garc\'{i}a-Berro}, E., {Isern}  \&
  {Burkert}}{{Garc\'{i}a-Berro} et~al.}{1999}]{GBerro1999}
{Garc\'{i}a-Berro} E. {Torres} S.,  {Isern} J.,   {Burkert} A.,  1999, \mnras,
  302, 173

\bibitem[\protect\citeauthoryear{{Garc{\'{\i}}a-Berro}, {Torres}, {Isern}  \&
  {Burkert}}{{Garc{\'{\i}}a-Berro} et~al.}{2004}]{GBerro2004}
{Garc{\'{\i}}a-Berro} E.,  {Torres} S.,  {Isern} J.,   {Burkert} A.,  2004,
  \mn@doi [\aap] {10.1051/0004-6361:20034541}, \href
  {http://adsabs.harvard.edu/abs/2004A%26A...418...53G} {418, 53}

\bibitem[\protect\citeauthoryear{{Garc{\'{\i}}a-Berro}
  et~al.,}{{Garc{\'{\i}}a-Berro} et~al.}{2010}]{GBerro2010}
{Garc{\'{\i}}a-Berro} E.,  et~al., 2010, \mn@doi [\nat] {10.1038/nature09045},
  \href {http://adsabs.harvard.edu/abs/2010Natur.465..194G} {465, 194}

\bibitem[\protect\citeauthoryear{{Garc{\'{\i}}a-Berro}, {Lor{\'e}n-Aguilar},
  {Torres}, {Althaus}  \& {Isern}}{{Garc{\'{\i}}a-Berro}
  et~al.}{2011}]{GBerro2011}
{Garc{\'{\i}}a-Berro} E.,  {Lor{\'e}n-Aguilar} P.,  {Torres} S.,  {Althaus}
  L.~G.,   {Isern} J.,  2011, \mn@doi [\jcap] {10.1088/1475-7516/2011/05/021},
  \href {http://adsabs.harvard.edu/abs/2011JCAP...05..021G} {5, 021}

\bibitem[\protect\citeauthoryear{{Garc{\'{\i}}a-Berro}
  et~al.,}{{Garc{\'{\i}}a-Berro} et~al.}{2012}]{GBerro2012}
{Garc{\'{\i}}a-Berro} E.,  et~al., 2012, \mn@doi [\apj]
  {10.1088/0004-637X/749/1/25}, \href
  {http://adsabs.harvard.edu/abs/2012ApJ...749...25G} {749, 25}

\bibitem[\protect\citeauthoryear{{Giammichele}, {Bergeron}  \&
  {Dufour}}{{Giammichele} et~al.}{2012a}]{Giammichele12}
{Giammichele} N.,  {Bergeron} P.,   {Dufour} P.,  2012a, \mn@doi [\apjs]
  {10.1088/0067-0049/199/2/29}, \href
  {http://adsabs.harvard.edu/abs/2012ApJS..199...29G} {199, 29}

\bibitem[\protect\citeauthoryear{{Giammichele}, {Bergeron}  \&
  {Dufour}}{{Giammichele} et~al.}{2012b}]{Giammichele2012}
{Giammichele} N.,  {Bergeron} P.,   {Dufour} P.,  2012b, \mn@doi [\apjs]
  {10.1088/0067-0049/199/2/29}, \href
  {http://adsabs.harvard.edu/abs/2012ApJS..199...29G} {199, 29}

\bibitem[\protect\citeauthoryear{{Girven}, {G{\"a}nsicke}, {Steeghs}  \&
  {Koester}}{{Girven} et~al.}{2011}]{Girvenetal11}
{Girven} J.,  {G{\"a}nsicke} B.~T.,  {Steeghs} D.,   {Koester} D.,  2011,
  \mn@doi [\mnras] {10.1111/j.1365-2966.2011.19337.x}, \href
  {http://adsabs.harvard.edu/abs/2011MNRAS.417.1210G} {417, 1210}

\bibitem[\protect\citeauthoryear{{Hambly}, {Miller}, {MacGillivray}, {Herd}  \&
  {Cormack}}{{Hambly} et~al.}{1998}]{Hambly1998}
{Hambly} N.~C.,  {Miller} L.,  {MacGillivray} H.~T.,  {Herd} J.~T.,   {Cormack}
  W.~A.,  1998, \mn@doi [\mnras] {10.1046/j.1365-8711.1998.01669.x}, \href
  {http://adsabs.harvard.edu/abs/1998MNRAS.298..897H} {298, 897}

\bibitem[\protect\citeauthoryear{{Hansen} et~al.,}{{Hansen}
  et~al.}{2013}]{Hansen2013}
{Hansen} B.~M.~S.,  et~al., 2013, \mn@doi [\nat] {10.1038/nature12334}, \href
  {http://adsabs.harvard.edu/abs/2013Natur.500...51H} {500, 51}

\bibitem[\protect\citeauthoryear{{Hewett}, {Warren}, {Leggett}  \&
  {Hodgkin}}{{Hewett} et~al.}{2006}]{Hewett06}
{Hewett} P.~C.,  {Warren} S.~J.,  {Leggett} S.~K.,   {Hodgkin} S.~T.,  2006,
  \mn@doi [\mnras] {10.1111/j.1365-2966.2005.09969.x}, \href
  {http://adsabs.harvard.edu/abs/2006MNRAS.367..454H} {367, 454}

\bibitem[\protect\citeauthoryear{{Holberg}, {Sion}, {Oswalt}, {McCook}, {Foran}
   \& {Subasavage}}{{Holberg} et~al.}{2008a}]{Holbergetal08}
{Holberg} J.~B.,  {Sion} E.~M.,  {Oswalt} T.,  {McCook} G.~P.,  {Foran} S.,
  {Subasavage} J.~P.,  2008a, \mn@doi [\aj] {10.1088/0004-6256/135/4/1225},
  \href {http://adsabs.harvard.edu/abs/2008AJ....135.1225H} {135, 1225}

\bibitem[\protect\citeauthoryear{{Holberg}, {Sion}, {Oswalt}, {McCook}, {Foran}
   \& {Subasavage}}{{Holberg} et~al.}{2008b}]{Holberg2008}
{Holberg} J.~B.,  {Sion} E.~M.,  {Oswalt} T.,  {McCook} G.~P.,  {Foran} S.,
  {Subasavage} J.~P.,  2008b, \mn@doi [\aj] {10.1088/0004-6256/135/4/1225},
  \href {http://adsabs.harvard.edu/abs/2008AJ....135.1225H} {135, 1225}

\bibitem[\protect\citeauthoryear{{Holberg}, {Oswalt}, {Sion}  \&
  {McCook}}{{Holberg} et~al.}{2016}]{Holberg2016}
{Holberg} J.~B.,  {Oswalt} T.~D.,  {Sion} E.~M.,   {McCook} G.~P.,  2016,
  \mn@doi [\mnras] {10.1093/mnras/stw1357}, \href
  {http://adsabs.harvard.edu/abs/2016MNRAS.462.2295H} {462, 2295}

\bibitem[\protect\citeauthoryear{{Isern}, {Hernanz}  \& {Garcia-Berro}}{{Isern}
  et~al.}{1992}]{Isern1992}
{Isern} J.,  {Hernanz} M.,   {Garcia-Berro} E.,  1992, \mn@doi [\apjl]
  {10.1086/186416}, \href {http://adsabs.harvard.edu/abs/1992ApJ...392L..23I}
  {392, L23}

\bibitem[\protect\citeauthoryear{{Isern}, {Garc{\'{\i}}a-Berro}, {Hernanz},
  {Mochkovitch}  \& {Torres}}{{Isern} et~al.}{1998}]{Isern1998}
{Isern} J.,  {Garc{\'{\i}}a-Berro} E.,  {Hernanz} M.,  {Mochkovitch} R.,
  {Torres} S.,  1998, \mn@doi [\apj] {10.1086/305977}, \href
  {http://adsabs.harvard.edu/abs/1998ApJ...503..239I} {503, 239}

\bibitem[\protect\citeauthoryear{{Isern}, {Garc{\'{\i}}a-Berro}, {Torres}  \&
  {Catal{\'a}n}}{{Isern} et~al.}{2008}]{Isern2008}
{Isern} J.,  {Garc{\'{\i}}a-Berro} E.,  {Torres} S.,   {Catal{\'a}n} S.,  2008,
  \mn@doi [\apjl] {10.1086/591042}, \href
  {http://adsabs.harvard.edu/abs/2008ApJ...682L.109I} {682, L109}

\bibitem[\protect\citeauthoryear{{James}}{{James}}{1990}]{James1990}
{James} F.,  1990, \mn@doi [Computer Physics Communications]
  {10.1016/0010-4655(90)90032-V}, \href
  {http://adsabs.harvard.edu/abs/1990CoPhC..60..329J} {60, 329}

\bibitem[\protect\citeauthoryear{{Jeffery}, {von Hippel}, {DeGennaro}, {van
  Dyk}, {Stein}  \& {Jefferys}}{{Jeffery} et~al.}{2011}]{Jeffery2011}
{Jeffery} E.~J.,  {von Hippel} T.,  {DeGennaro} S.,  {van Dyk} D.~A.,  {Stein}
  N.,   {Jefferys} W.~H.,  2011, \mn@doi [\apj] {10.1088/0004-637X/730/1/35},
  \href {http://adsabs.harvard.edu/abs/2011ApJ...730...35J} {730, 35}

\bibitem[\protect\citeauthoryear{{Jordi} et~al.,}{{Jordi}
  et~al.}{2010}]{Jordi2010}
{Jordi} C.,  et~al., 2010, \mn@doi [\aap] {10.1051/0004-6361/201015441}, \href
  {http://adsabs.harvard.edu/abs/2010A%26A...523A..48J} {523, A48}

\bibitem[\protect\citeauthoryear{{Kalirai}, {Hansen}, {Kelson}, {Reitzel},
  {Rich}  \& {Richer}}{{Kalirai} et~al.}{2008}]{Kalirai2008}
{Kalirai} J.~S.,  {Hansen} B.~M.~S.,  {Kelson} D.~D.,  {Reitzel} D.~B.,  {Rich}
  R.~M.,   {Richer} H.~B.,  2008, \mn@doi [\apj] {10.1086/527028}, \href
  {http://adsabs.harvard.edu/abs/2008ApJ...676..594K} {676, 594}

\bibitem[\protect\citeauthoryear{{Kepler} et~al.,}{{Kepler}
  et~al.}{2015}]{Kepleretal15}
{Kepler} S.~O.,  et~al., 2015, \mn@doi [\mnras] {10.1093/mnras/stu2388}, \href
  {http://adsabs.harvard.edu/abs/2015MNRAS.446.4078K} {446, 4078}

\bibitem[\protect\citeauthoryear{{Kepler} et~al.,}{{Kepler}
  et~al.}{2016}]{Kepler16}
{Kepler} S.~O.,  et~al., 2016, \mn@doi [\mnras] {10.1093/mnras/stv2526}, \href
  {http://adsabs.harvard.edu/abs/2016MNRAS.455.3413K} {455, 3413}

\bibitem[\protect\citeauthoryear{{Kilic}, {Brown}, {Allende Prieto}, {Kenyon},
  {Heinke}, {Ag{\"u}eros}  \& {Kleinman}}{{Kilic} et~al.}{2012}]{Kilicetal12}
{Kilic} M.,  {Brown} W.~R.,  {Allende Prieto} C.,  {Kenyon} S.~J.,  {Heinke}
  C.~O.,  {Ag{\"u}eros} M.~A.,   {Kleinman} S.~J.,  2012, \mn@doi [\apj]
  {10.1088/0004-637X/751/2/141}, \href
  {http://adsabs.harvard.edu/abs/2012ApJ...751..141K} {751, 141}

\bibitem[\protect\citeauthoryear{{Kilic}, {Hambly}, {Bergeron},
  {Genest-Beaulieu}  \& {Rowell}}{{Kilic} et~al.}{2018}]{Kilic2018}
{Kilic} M.,  {Hambly} N.~C.,  {Bergeron} P.,  {Genest-Beaulieu} C.,   {Rowell}
  N.,  2018, \mn@doi [\mnras] {10.1093/mnrasl/sly110}, \href
  {http://adsabs.harvard.edu/abs/2018MNRAS.tmpL.114K} {}

\bibitem[\protect\citeauthoryear{{Kleinman} et~al.,}{{Kleinman}
  et~al.}{2013}]{Kleinmanetal13}
{Kleinman} S.~J.,  et~al., 2013, \mn@doi [\apjs] {10.1088/0067-0049/204/1/5},
  \href {http://cdsads.u-strasbg.fr/abs/2013ApJS..204....5K} {204, 5}

\bibitem[\protect\citeauthoryear{{Koester}}{{Koester}}{2010}]{Koester10}
{Koester} D.,  2010, \memsai, \href
  {http://adsabs.harvard.edu/abs/2010MmSAI..81..921K} {81, 921}

\bibitem[\protect\citeauthoryear{{Koester} \& {Kepler}}{{Koester} \&
  {Kepler}}{2015}]{Koester2015}
{Koester} D.,  {Kepler} S.~O.,  2015, \mn@doi [\aap]
  {10.1051/0004-6361/201527169}, \href
  {http://adsabs.harvard.edu/abs/2015A%26A...583A..86K} {583, A86}

\bibitem[\protect\citeauthoryear{{Koester}, {Schulz}  \& {Weidemann}}{{Koester}
  et~al.}{1979}]{Koesteretal79}
{Koester} D.,  {Schulz} H.,   {Weidemann} V.,  1979, \aap, 76, 262

\bibitem[\protect\citeauthoryear{{Liebert}, {Bergeron}  \& {Holberg}}{{Liebert}
  et~al.}{2005}]{Liebert05}
{Liebert} J.,  {Bergeron} P.,   {Holberg} J.~B.,  2005, apjss, \href
  {2005ApJS..156...47L} {156, 47}

\bibitem[\protect\citeauthoryear{{Limoges}, {Bergeron}  \&
  {L{\'e}pine}}{{Limoges} et~al.}{2015}]{Limoges15}
{Limoges} M.-M.,  {Bergeron} P.,   {L{\'e}pine} S.,  2015, \mn@doi [\apjs]
  {10.1088/0067-0049/219/2/19}, \href
  {http://adsabs.harvard.edu/abs/2015ApJS..219...19L} {219, 19}

\bibitem[\protect\citeauthoryear{{Lindegren} et~al.,}{{Lindegren}
  et~al.}{2018}]{Lindegren18}
{Lindegren} L.,  et~al., 2018, preprint, \href
  {http://cdsads.u-strasbg.fr/abs/2018arXiv180409366L} {} (\mn@eprint {arXiv}
  {1804.09366})

\bibitem[\protect\citeauthoryear{{Lutz} \& {Kelker}}{{Lutz} \&
  {Kelker}}{1973}]{LutzKelker1973}
{Lutz} T.~E.,  {Kelker} D.~H.,  1973, \mn@doi [\pasp] {10.1086/129506}, \href
  {http://adsabs.harvard.edu/abs/1973PASP...85..573L} {85, 573}

\bibitem[\protect\citeauthoryear{{Maoz}, {Hallakoun}  \& {Badenes}}{{Maoz}
  et~al.}{2018}]{Maoz18}
{Maoz} D.,  {Hallakoun} N.,   {Badenes} C.,  2018, \mn@doi [\mnras]
  {10.1093/mnras/sty339}, \href
  {http://adsabs.harvard.edu/abs/2018MNRAS.476.2584M} {476, 2584}

\bibitem[\protect\citeauthoryear{{Meng}, {Chen}  \& {Han}}{{Meng}
  et~al.}{2008}]{Meng2008}
{Meng} X.,  {Chen} X.,   {Han} Z.,  2008, \mn@doi [\aap]
  {10.1051/0004-6361:20078841}, \href
  {http://adsabs.harvard.edu/abs/2008A%26A...487..625M} {487, 625}

\bibitem[\protect\citeauthoryear{{Miller Bertolami}, {Melendez}, {Althaus}  \&
  {Isern}}{{Miller Bertolami} et~al.}{2014}]{Miller2014}
{Miller Bertolami} M.~M.,  {Melendez} B.~E.,  {Althaus} L.~G.,   {Isern} J.,
  2014, \mn@doi [\jcap] {10.1088/1475-7516/2014/10/069}, \href
  {http://adsabs.harvard.edu/abs/2014JCAP...10..069M} {10, 069}

\bibitem[\protect\citeauthoryear{{Mochkovitch}, {Garcia-Berro}, {Hernanz},
  {Isern}  \& {Panis}}{{Mochkovitch} et~al.}{1990}]{Mochkovitch1990}
{Mochkovitch} R.,  {Garcia-Berro} E.,  {Hernanz} M.,  {Isern} J.,   {Panis}
  J.~F.,  1990, \aap, 233, 456

\bibitem[\protect\citeauthoryear{{Moles} et~al.,}{{Moles}
  et~al.}{2008}]{Moles08}
{Moles} M.,  et~al., 2008, \mn@doi [\aj] {10.1088/0004-6256/136/3/1325}, \href
  {http://adsabs.harvard.edu/abs/2008AJ....136.1325M} {136, 1325}

\bibitem[\protect\citeauthoryear{{Parsons} et~al.,}{{Parsons}
  et~al.}{2017}]{Parsons2017}
{Parsons} S.~G.,  et~al., 2017, \mn@doi [\mnras] {10.1093/mnras/stx1522}, \href
  {http://adsabs.harvard.edu/abs/2017MNRAS.470.4473P} {470, 4473}

\bibitem[\protect\citeauthoryear{{Provencal}, {Shipman}, {H{\o}g}  \&
  {Thejll}}{{Provencal} et~al.}{1998}]{Provencal1998}
{Provencal} J.~L.,  {Shipman} H.~L.,  {H{\o}g} E.,   {Thejll} P.,  1998,
  \mn@doi [\apj] {10.1086/305238}, \href
  {http://adsabs.harvard.edu/abs/1998ApJ...494..759P} {494, 759}

\bibitem[\protect\citeauthoryear{{Rebassa-Mansergas}, {G{\"a}nsicke},
  {Rodr{\'{\i}}guez-Gil}, {Schreiber}  \& {Koester}}{{Rebassa-Mansergas}
  et~al.}{2007}]{Rebassa07}
{Rebassa-Mansergas} A.,  {G{\"a}nsicke} B.~T.,  {Rodr{\'{\i}}guez-Gil} P.,
  {Schreiber} M.~R.,   {Koester} D.,  2007, \mn@doi [\mnras]
  {10.1111/j.1365-2966.2007.12288.x}, \href {2007MNRAS.382.1377R} {382, 1377}

\bibitem[\protect\citeauthoryear{{Rebassa-Mansergas}, {Nebot
  G{\'o}mez-Mor{\'a}n}, {Schreiber}, {Girven}  \&
  {G{\"a}nsicke}}{{Rebassa-Mansergas} et~al.}{2011}]{Rebassa11}
{Rebassa-Mansergas} A.,  {Nebot G{\'o}mez-Mor{\'a}n} A.,  {Schreiber} M.~R.,
  {Girven} J.,   {G{\"a}nsicke} B.~T.,  2011, \mn@doi [\mnras]
  {10.1111/j.1365-2966.2011.18200.x}, \href
  {http://adsabs.harvard.edu/abs/2011MNRAS.413.1121R} {413, 1121}

\bibitem[\protect\citeauthoryear{{Rebassa-Mansergas}
  et~al.,}{{Rebassa-Mansergas} et~al.}{2015a}]{Rebassa15-1}
{Rebassa-Mansergas} A.,  et~al., 2015a, \mn@doi [\mnras]
  {10.1093/mnras/stv607}, \href
  {http://adsabs.harvard.edu/abs/2015MNRAS.450..743R} {450, 743}

\bibitem[\protect\citeauthoryear{{Rebassa-Mansergas}, {Rybicka}, {Liu}, {Han}
  \& {Garc{\'{\i}}a-Berro}}{{Rebassa-Mansergas} et~al.}{2015b}]{Rebassa15-2}
{Rebassa-Mansergas} A.,  {Rybicka} M.,  {Liu} X.-W.,  {Han} Z.,
  {Garc{\'{\i}}a-Berro} E.,  2015b, \mn@doi [\mnras] {10.1093/mnras/stv1399},
  \href {http://adsabs.harvard.edu/abs/2015MNRAS.452.1637R} {452, 1637}

\bibitem[\protect\citeauthoryear{{Renedo}, {Althaus}, {Miller Bertolami},
  {Romero}, {C{\'o}rsico}, {Rohrmann}  \& {Garc{\'{\i}}a-Berro}}{{Renedo}
  et~al.}{2010}]{Renedo2010}
{Renedo} I.,  {Althaus} L.~G.,  {Miller Bertolami} M.~M.,  {Romero} A.~D.,
  {C{\'o}rsico} A.~H.,  {Rohrmann} R.~D.,   {Garc{\'{\i}}a-Berro} E.,  2010,
  \apj, 717, 183

\bibitem[\protect\citeauthoryear{{Rohrmann}}{{Rohrmann}}{2001}]{Rohrmann2001}
{Rohrmann} R.~D.,  2001, \mn@doi [\mnras] {10.1046/j.1365-8711.2001.04298.x},
  \href {http://adsabs.harvard.edu/abs/2001MNRAS.323..699R} {323, 699}

\bibitem[\protect\citeauthoryear{{Rowell}}{{Rowell}}{2013}]{Rowell2013}
{Rowell} N.,  2013, \mn@doi [\mnras] {10.1093/mnras/stt1110}, \href
  {http://adsabs.harvard.edu/abs/2013MNRAS.434.1549R} {434, 1549}

\bibitem[\protect\citeauthoryear{{Rowell} \& {Hambly}}{{Rowell} \&
  {Hambly}}{2011}]{Rowell2011}
{Rowell} N.,  {Hambly} N.~C.,  2011, \mn@doi [\mnras]
  {10.1111/j.1365-2966.2011.18976.x}, \href
  {http://cdsads.u-strasbg.fr/abs/2011MNRAS.417...93R} {417, 93}

\bibitem[\protect\citeauthoryear{{Salaris}, {Cassisi}, {Garc{\'{\i}}a-Berro},
  {Isern}  \& {Torres}}{{Salaris} et~al.}{2001}]{Salaris2001}
{Salaris} M.,  {Cassisi} S.,  {Garc{\'{\i}}a-Berro} E.,  {Isern} J.,   {Torres}
  S.,  2001, \mn@doi [\aap] {10.1051/0004-6361:20010388}, \href
  {http://adsabs.harvard.edu/abs/2001A%26A...371..921S} {371, 921}

\bibitem[\protect\citeauthoryear{{Salpeter}}{{Salpeter}}{1955}]{Salpeter1955}
{Salpeter} E.~E.,  1955, \mn@doi [\apj] {10.1086/145971}, \href
  {http://adsabs.harvard.edu/abs/1955ApJ...121..161S} {121, 161}

\bibitem[\protect\citeauthoryear{{Schlafly} \& {Finkbeiner}}{{Schlafly} \&
  {Finkbeiner}}{2011}]{Schlafly+Finkbeiner11}
{Schlafly} E.~F.,  {Finkbeiner} D.~P.,  2011, \mn@doi [\apj]
  {10.1088/0004-637X/737/2/103}, \href
  {http://adsabs.harvard.edu/abs/2011ApJ...737..103S} {737, 103}

\bibitem[\protect\citeauthoryear{{Serenelli}, {Althaus}, {Rohrmann}  \&
  {Benvenuto}}{{Serenelli} et~al.}{2001}]{Serenelli2001}
{Serenelli} A.~M.,  {Althaus} L.~G.,  {Rohrmann} R.~D.,   {Benvenuto} O.~G.,
  2001, \mn@doi [\mnras] {10.1046/j.1365-8711.2001.04449.x}, \href
  {http://adsabs.harvard.edu/abs/2001MNRAS.325..607S} {325, 607}

\bibitem[\protect\citeauthoryear{{Skrutskie} et~al.,}{{Skrutskie}
  et~al.}{2006}]{Skrutskie06}
{Skrutskie} M.~F.,  et~al., 2006, \mn@doi [\aj] {10.1086/498708}, \href
  {http://adsabs.harvard.edu/abs/2006AJ....131.1163S} {131, 1163}

\bibitem[\protect\citeauthoryear{{Taylor}}{{Taylor}}{2005}]{Taylor05}
{Taylor} M.~B.,  2005, in {Shopbell} P.,  {Britton} M.,   {Ebert} R.,  eds,
  Astronomical Society of the Pacific Conference Series Vol. 347, Astronomical
  Data Analysis Software and Systems XIV. p.~29

\bibitem[\protect\citeauthoryear{{Torres} \& {Garc{\'{\i}}a-Berro}}{{Torres} \&
  {Garc{\'{\i}}a-Berro}}{2016}]{Torres2016}
{Torres} S.,  {Garc{\'{\i}}a-Berro} E.,  2016, \mn@doi [\aap]
  {10.1051/0004-6361/201528059}, \href
  {http://adsabs.harvard.edu/abs/2016A%26A...588A..35T} {588, A35}

\bibitem[\protect\citeauthoryear{Torres, Garc{\'{\i}}a-Berro, Burkert  \&
  Isern}{Torres et~al.}{2001}]{Torres2001}
Torres S.,  Garc{\'{\i}}a-Berro E.,  Burkert A.,   Isern J.,  2001, \mn@doi
  [\mnras] {10.1046/j.1365-8711.2001.04885.x}, 328, 492

\bibitem[\protect\citeauthoryear{{Torres}, {Garc{\'{\i}}a-Berro}, {Burkert}  \&
  {Isern}}{{Torres} et~al.}{2002}]{Torres2002}
{Torres} S.,  {Garc{\'{\i}}a-Berro} E.,  {Burkert} A.,   {Isern} J.,  2002,
  \mn@doi [\mnras] {10.1046/j.1365-8711.2002.05830.x}, \href
  {http://adsabs.harvard.edu/abs/2002MNRAS.336..971T} {336, 971}

\bibitem[\protect\citeauthoryear{{Torres}, {Garc{\'{\i}}a-Berro}, {Isern}  \&
  {Figueras}}{{Torres} et~al.}{2005}]{Torres2005}
{Torres} S.,  {Garc{\'{\i}}a-Berro} E.,  {Isern} J.,   {Figueras} F.,  2005,
  \mn@doi [\mnras] {10.1111/j.1365-2966.2005.09128.x}, \href
  {http://adsabs.harvard.edu/abs/2005MNRAS.360.1381T} {360, 1381}

\bibitem[\protect\citeauthoryear{{Torres}, {Garc{\'{\i}}a-Berro}  \&
  {Isern}}{{Torres} et~al.}{2007}]{Torres2007}
{Torres} S.,  {Garc{\'{\i}}a-Berro} E.,   {Isern} J.,  2007, \mn@doi [\mnras]
  {10.1111/j.1365-2966.2007.11887.x}, \href
  {http://adsabs.harvard.edu/abs/2007MNRAS.378.1461T} {378, 1461}

\bibitem[\protect\citeauthoryear{{Torres}, {Garc{\'{\i}}a-Berro}, {Althaus}  \&
  {Camisassa}}{{Torres} et~al.}{2015}]{Torres2015}
{Torres} S.,  {Garc{\'{\i}}a-Berro} E.,  {Althaus} L.~G.,   {Camisassa} M.~E.,
  2015, \mn@doi [\aap] {10.1051/0004-6361/201526157}, \href
  {http://adsabs.harvard.edu/abs/2015A%26A...581A..90T} {581, A90}

\bibitem[\protect\citeauthoryear{{Torres}, {Garc{\'{\i}}a-Berro}, {Cojocaru}
  \& {Calamida}}{{Torres} et~al.}{2018}]{Torres2018}
{Torres} S.,  {Garc{\'{\i}}a-Berro} E.,  {Cojocaru} R.,   {Calamida} A.,  2018,
  \mn@doi [\mnras] {10.1093/mnras/sty289}, \href
  {http://adsabs.harvard.edu/abs/2018MNRAS.tmp..283T} {}

\bibitem[\protect\citeauthoryear{{Tremblay} \& {Bergeron}}{{Tremblay} \&
  {Bergeron}}{2008}]{Tremblay2008}
{Tremblay} P.-E.,  {Bergeron} P.,  2008, \mn@doi [\apj] {10.1086/524134}, \href
  {http://adsabs.harvard.edu/abs/2008ApJ...672.1144T} {672, 1144}

\bibitem[\protect\citeauthoryear{{Weidemann}}{{Weidemann}}{1968}]{weid68}
{Weidemann} V.,  1968, \mn@doi [\araa] {10.1146/annurev.aa.06.090168.002031},
  \href {http://adsabs.harvard.edu/abs/1968ARA%26A...6..351W} {6, 351}

\bibitem[\protect\citeauthoryear{{Williams}, {Bolte}  \& {Koester}}{{Williams}
  et~al.}{2004}]{Williams2004}
{Williams} K.~A.,  {Bolte} M.,   {Koester} D.,  2004, \mn@doi [\apjl]
  {10.1086/425995}, \href {http://adsabs.harvard.edu/abs/2004ApJ...615L..49W}
  {615, L49}

\bibitem[\protect\citeauthoryear{{Winget}, {Hansen}, {Liebert}, {van Horn},
  {Fontaine}, {Nather}, {Kepler}  \& {Lamb}}{{Winget} et~al.}{1987}]{Winget87}
{Winget} D.~E.,  {Hansen} C.~J.,  {Liebert} J.,  {van Horn} H.~M.,  {Fontaine}
  G.,  {Nather} R.~E.,  {Kepler} S.~O.,   {Lamb} D.~Q.,  1987, \mn@doi [\apjl]
  {10.1086/184864}, \href {http://adsabs.harvard.edu/abs/1987ApJ...315L..77W}
  {315, L77}

\bibitem[\protect\citeauthoryear{{York} et~al.,}{{York}
  et~al.}{2000}]{Yorketal2000}
{York} D.~G.,  et~al., 2000, \mn@doi [\aj] {10.1086/301513}, \href
  {http://adsabs.harvard.edu/abs/2000AJ....120.1579Y} {120, 1579}

\bibitem[\protect\citeauthoryear{{de~Bruijne}, Perryman, Lindegren
  et~al.}{{de~Bruijne} et~al.}{2005}]{Bruijne2005}
{de~Bruijne} J.,  Perryman M.,  Lindegren L.,   et~al., 2005, {G}aia
  astrometric, photometric, and radial-velocity performance assessment
  methodologies to be used by the industrial system-level teams, GAIA-JDB-022,
  \url {http://www.rssd.esa.int/doc_fetch.php?id=448635}

\bibitem[\protect\citeauthoryear{{van Oirschot}, {Nelemans}, {Toonen}, {Pols},
  {Brown}, {Helmi}  \& {Portegies Zwart}}{{van Oirschot}
  et~al.}{2014}]{vanOirschot2014}
{van Oirschot} P.,  {Nelemans} G.,  {Toonen} S.,  {Pols} O.,  {Brown} A.~G.~A.,
   {Helmi} A.,   {Portegies Zwart} S.,  2014, \aap, 569, A42

\makeatother
\end{thebibliography}


\appendix

\section{Online catalogue service}
\label{append}

In order to help the astronomical community on using our catalogue of white dwarfs, we developed a wed archive system that can be accessed from a webpage\footnote{http://svo2.cab.inta-csic.es/vocats/v2/wdw/} or through a Virtual Observatory ConeSearch\footnote{e.g. http://svo2.cab.inta-csic.es/vocats/v2/wdw/cs.php?RA=301.708\&DEC=-67.482\&SR=0.1\&VERB=2}.

The archive system implements a very simple search interface that permits queries by coordinates and radius as well as by other parameters of interest. The user can also select the maximum number of sources (with values from 10 to unlimited) and the number of columns to return (minimum, default, or maximum verbosity).

The result of the query is a HTML table with all the sources found in the archive fulfilling the search criteria. The result can also be downloaded as a VOTable or a CSV file. Detailed information on the output fields can be obtained placing the mouse over the question mark (``?") located close to the name of the column. The archive also implements the SAMP\footnote{http://www.ivoa.net/documents/SAMP/} (Simple Application Messaging) Virtual Observatory protocol. SAMP allows Virtual Observatory applications to communicate with each other in a seamless and transparent manner for the user. This way, the results of a query can be easily transferred to other VO applications, such as, for instance, Topcat.



\bsp	
\label{lastpage}
\end{document}